\documentclass[12pt]{article}

\usepackage{hyperref}
\usepackage{url}
\usepackage{booktabs}
\usepackage{amsfonts, amsmath, amsthm, amssymb}
\usepackage{nicefrac}
\usepackage{microtype}
\usepackage{bm}
\usepackage{mathabx}
\usepackage{pdfpages}
\usepackage{enumerate}
\usepackage{makecell}
\usepackage{subcaption}
\usepackage{natbib}
\usepackage{algpseudocode,algorithm}
\usepackage{caption}
\newcommand{\srs}{\hspace{6pt}\Rightarrow\hspace{6pt}}
\newcommand{\sls}{\hspace{6pt}\Leftarrow\hspace{6pt}}
\newcommand{\slrs}{\hspace{6pt}\Leftrightarrow\hspace{6pt}}

\newcommand{\bZ}{\bm{Z}}
\newcommand{\mP}{\mathcal{P}}
\newcommand{\bX}{\mathbf{X}}
\newcommand{\bY}{\mathbf{Y}}
\newcommand{\bk}{\mathbf{k}}
\newcommand{\bl}{\mathbf{l}}
\newcommand{\bn}{\mathbf{n}}
\newcommand{\N}{\mathbb{N}}

\newcommand{\D}{\{1,..,D\}}
\newcommand{\mA}{\mathcal{A}}
\newcommand{\Pb}{\operatorname{P}}
\renewcommand{\P}{\mathcal{P}}

\newcommand{\Dx}{\{1,..,D_x\}}
\newcommand{\Dyd}{\{D_x+1,..,D\}}
\newcommand{\Dy}{\{1,..,D_y\}}
\newcommand{\bi}{\mathbf{i}}
\newcommand{\bj}{\mathbf{j}}
\newcommand{\bd}{\mathbf{d}}
\newcommand{\FX}{F_{\bX}}
\newcommand{\FY}{F_{\bY}}
\newcommand\C{\mathcal{C}}
\newcommand{\A}{\mathcal{A}}
\newcommand{\om}{\Omega}
\newcommand\indep{\protect\mathpalette{\protect\independenT}{\perp}}
\def\independenT#1#2{\mathrel{\rlap{$#1#2$}\mkern2mu{#1#2}}}

\newcommand\indepk{\protect\mathpalette{\protect\independenTk}{\perp}}
\def\independenTk#1#2{\mathrel{\rlap{$#1#2$}\mkern2mu{#1#2}}_\bk}

\newcommand\indepkp{\protect\mathpalette{\protect\independenTkp}{\perp}}
\def\independenTkp#1#2{\mathrel{\rlap{$#1#2$}\mkern2mu{#1#2}}_{\bk'}}

\newcommand\indepkd{\protect\mathpalette{\protect\independenTkd}{\perp}}
\def\independenTkd#1#2{\mathrel{\rlap{$#1#2$}\mkern2mu{#1#2}}_{\bk_{(\bd)}}}

\newcommand\indepkdd{\protect\mathpalette{\protect\independenTkdd}{\perp}}
\def\independenTkdd#1#2{\mathrel{\rlap{$#1#2$}\mkern2mu{#1#2}}_{\bk_{\bd}}}

\newcommand\indepkij{\protect\mathpalette{\protect\independenTkij}{\perp}}
\def\independenTkij#1#2{\mathrel{\rlap{$#1#2$}\mkern2mu{#1#2}}_{(k_i,k_{D_x+j})}}

\newcommand{\beginsupplement}{%
        \setcounter{section}{0}
        \renewcommand{\thesection}{S\arabic{section}}%
        \setcounter{table}{0}
        \renewcommand{\thetable}{S\arabic{table}}%
        \setcounter{figure}{0}
        \renewcommand{\thefigure}{S\arabic{figure}}%
        \setcounter{definition}{0}
        \renewcommand{\thedefinition}{S\arabic{section}.\arabic{definition}}%
        \setcounter{lemma}{0}
        \renewcommand{\thelemma}{S\arabic{section}.\arabic{lemma}}%
        \setcounter{corollary}{0}
        \renewcommand{\thecorollary}{S\arabic{section}.\arabic{corollary}}%
        \setcounter{theorem}{0}
        \renewcommand{\thetheorem}{S\arabic{section}.\arabic{theorem}}%
     }
     
\theoremstyle{plain}
\newtheorem{theorem}{Theorem}[section]

\theoremstyle{definition}
\newtheorem{definition}{Definition}[section]

\theoremstyle{plain}
\newtheorem{lemma}{Lemma}[section]

\theoremstyle{plain}
\newtheorem{corollary}{Corollary}[section]

\theoremstyle{plain}

\newcommand{\blind}{0}

\begin{document}

\def\spacingset#1{\renewcommand{\baselinestretch}%
{#1}\small\normalsize} \spacingset{1}

\if0\blind
{
  \title{\bf Multiscale Fisher's Independence Test for Multivariate Dependence}
  \author{Shai Gorsky\\
    Department of Mathematics and Statistics,\\University of Massachusetts, Amherst\\
    \\
    Li Ma
    \\
    Department of Statistical Science,\\Duke University}
  \maketitle
} \fi

\if1\blind
{
  \bigskip
  \bigskip
  \begin{center}
    {\LARGE\bf MultiFIT: Multiscale Fisher's Independence Test for Multivariate Dependence}
\end{center}
} \fi

\begin{abstract}
Identifying dependency in multivariate data is a common inference task that arises in numerous applications. However, existing nonparametric independence tests typically require computation that scales at least quadratically with the sample size, making it difficult to apply them in the presence of massive sample sizes. Moreover, resampling (e.g., permutation) is usually necessary to evaluate the statistical significance of the resulting test statistics at finite sample sizes, further worsening the computational burden. We introduce a scalable, resampling-free approach to testing the independence between two random vectors by breaking down the task into simple univariate tests of independence on a collection of $2\times 2$ contingency tables constructed through sequential coarse-to-fine discretization of the sample space, transforming the inference task into a multiple testing problem that can be completed with almost linear complexity with respect to the sample size. To address increasing dimensionality, we introduce a coarse-to-fine sequential adaptive procedure that exploits the spatial features of dependency structures. We derive a finite-sample theory that guarantees the inferential validity of our adaptive procedure at any given sample size. We show that our approach can achieve strong control of the level of the testing procedure at any sample size without resampling or asymptotic approximation and establish its large-sample consistency. We demonstrate through an extensive simulation study its substantial computational advantage in comparison to existing approaches while achieving robust statistical power under various dependency scenarios, and illustrate how the divide-and-conquer nature can be exploited to not just test independence but to learn the nature of the underlying dependency. Finally, we demonstrate the use of our method through analyzing a dataset from a flow cytometry experiment.
\end{abstract}

\noindent%
\vfill

\newpage
\spacingset{1.45}

\section{Introduction}
\label{sec:intro}
Testing independence and learning the dependency structure in multivariate problems has been a central inference task since the very beginning of modern statistics, and the last two decades have witnessed a surge of interest in this problem among statisticians, engineers, and computer scientists. A variety of different methods have been proposed for testing independence between two random vectors. For example, \cite{szekely2009} generalize the product-moment covariance and correlation to the distance covariance and correlation. 
\cite{bakirov2006}, \cite{fan2017}, and \cite{meintanis2008} all developed nonparametric tests of independence based on the distance between the empirical joint characteristic function of the random vectors and the product of the marginal empirical characteristic functions of the two random vectors. \cite{szekely2013ttest} further consider an asymptotic scenario with the dimensionality of the vectors increasing to infinity while keeping the sample size fixed. 
In a different vein, \cite{heller2013} form a test based on univariate tests of independence between the distances of each of the random vectors from a central point. In machine learning, a class of kernel-based tests has also become popular. For example, \cite{gretton2007} form a test based on the eigenspectrum of covariance operators in a reproducing kernel Hilbert spaces (RKHS). More recently, \cite{pfister2018} generalized this approach to the multivariate case by embedding the joint distribution into an RKHS. \cite{weihs2018} defined a class of multivariate nonparametric measures which leads to multivariate extensions of the Bergsma-Dassios sign covariance. \cite{Lee2019} proposed using random projections to reduce multivariate independence testing to a univariate problem, and complete the latter using an ensemble approach combining the distance correlation and a binary expansion test statistic \citep{zhang2017}. 

The existing multivariate independence tests generally require the computation of statistics at a computational complexity that scales at least quadratically in the sample size, making them impractical for data sets with sample sizes greater than, say, tens of thousands of observations. Many of these multivariate methods also require resampling -- in the form of either permutation or bootstrap -- to evaluate statistical significance. This additional computational burden makes applications of these methods  computationally expensive even for problems with moderate sample sizes. To overcome these challenges, some appeal to asymptotic approximations (either in large $n$ or in large $p$) \citep{szekely2013ttest,pfister2018} to derive procedures that when the asymptotic conditions are satisfied do not require resampling. However, because it is hard to judge whether such conditions are true in multivariate settings, 
 practitioners usually still resort to resampling to ensure validity. 

A scalable testing strategy for data with massive sample sizes should ideally achieve (i) close to linear computational complexity in the sample size and (ii) finite-sample guarantees without the need for resampling or asymptotic approximation. 
We aim to introduce a framework that achieves these two desiderata. Specifically, instead of calculating a single test statistic for independence all at once, we take a multi-scale divide-and-conquer approach that breaks apart the nonparametric multivariate test of independence into simple univariate independence tests on a collection of $2\times 2$ contingency tables defined by sequentially discretizing the original sample space at a cascade of scales. This approach transforms a complex nonparametric testing problem into a multiple testing problem involving simple tests that can be carried out efficiently. 
While such an approach was previously adopted in \cite{mamao2017} for testing the independence between two scalar variables, the increasing dimensionality in the multivariate setting makes a brute-force, exhaustive approach as proposed therein computationally prohibitive and statistically inefficient. As such we incorporate data-adaptivity into the framework and introduce a coarse-to-fine sequential adaptive testing procedure which exploits the spatial characteristics of dependency structures to drastically reduce the number of univariate tests completed in the procedure. 
At the same time, we derive a finite-sample theory showing that even with the additional adaptivity, exact inference (in terms of controlling the level of the test) can be achieved at any given sample size without resorting to either resampling or large-sample approximation.

Aside from these properties, our approach also enjoys a unique feature of practical relevance---its divide-and-conquer nature allows learning the structure of the underlying dependency. 
By identifying and visualizing the $2\times 2$ tables on which the univariate independence test returns the most significant $p$-values, we can identify using {\tt MultiFIT} interesting dependency relationships otherwise hidden by the multivariate nature of the sample space and the complexity of the joint distribution.

We carry out extensive simulation studies that examine the computational scalability and statistical power of our method in a variety of dependency scenarios and compare our method to a number of state-of-the-art approaches. 
We demonstrate an application of our method to a data set from a flow cytometry experiment with a massive sample size. 
All technical proofs are provided in the Online Supplementary Materials~\ref{sec:supplement-proofs}.

\section{Method}
\label{sec:method}
Our strategy is to transform nonparametric testing of multivariate independence into a multiple testing problem involving univariate independence tests on a collection of $2\times 2$ tables constructed by sequentially partitioning the sample space. In Section~\ref{sec:mvar}, we start by describing the construction of these $2\times 2$ tables and justify the testing strategy by showing that two random vectors are independent if and only if univariate independence holds on all of the $2\times 2$ tables so constructed. In Section~\ref{sec:algo}, we present a data-adaptive sequential testing procedure that completes the univariate tests on only a subset of the $2\times 2$ tables to accommodate increasing dimensionality of the random vectors. Finally in Section~\ref{sec:theory} we derive a finite-sample theory that provides guarantees for the validity of our procedure at any sample size without appealing to resampling or asymptotic approximations, and establish the large-sample consistency for our procedure.

\subsection{Multi-scale $2\times 2$ testing for multivariate independence}
\label{sec:mvar}
We start by introducing some notations that will be used throughout the paper as well as some concepts related to nested dyadic partitioning (NDP), which will be used for constructing the $2\times 2$ tables on which univariate independence tests are completed.

Let $\om=\om_{\bX}\times \om_{\bY}$ denote a $D$-dimensional joint sample space of two random vectors $\bX$ and $\bY$ where $\om_{\bX}$ and $\om_{\bY}$ are respectively the marginal sample space of $\bX$ and $\bY$. For simplicity, we assume that $\om_{\bX}=[0,1]^{D_x}$ and $\om_{\bY}=[0,1]^{D_y}$ -- that is, each marginal random variable of the two random vectors is supported on $[0,1]$. This costs no generality as other random variables can be mapped onto the unit interval through a CDF transform.

\begin{figure}[!h]
\centering
\includegraphics[width=0.5\textwidth]{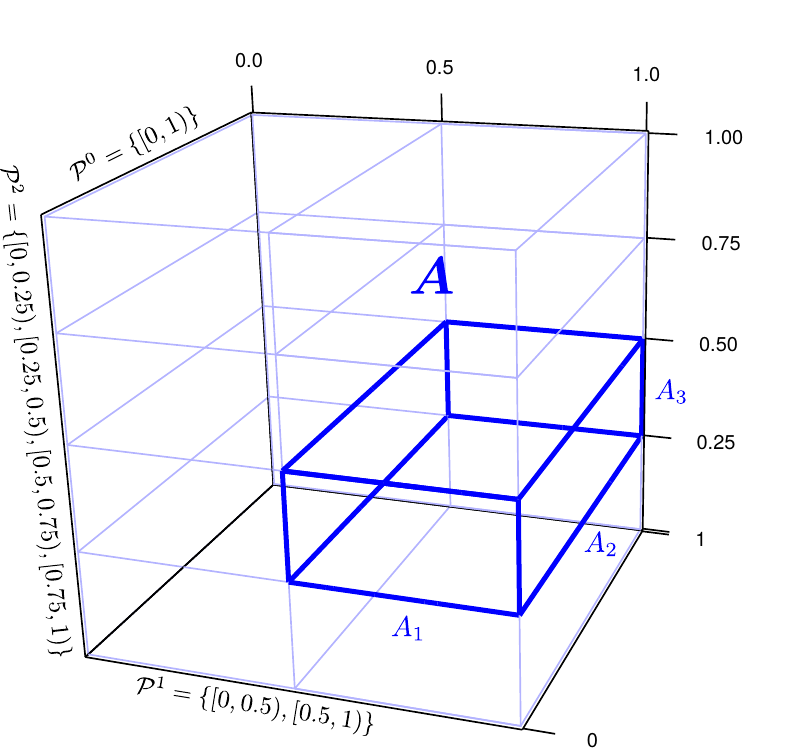}
\caption{\label{fig:cuboid-strat} 
A cuboid of resolution 3 in a 3-dimensional sample space under the canonical NDP.
}
\end{figure}

A {\em partition} $\P$ on a set $S$ is a collection of disjoint non-empty subsets of $S$ whose union is $S$. 
A {\em nested dyadic partition} (NDP) on $S$ is a sequence of partitions, $\P^0,\P^1,\ldots,\P^k,\ldots$ such that $\P^0=\{S\}$, and for each $k\geq 1$, the sets in $\P^{k}$ are those generated by dividing each set in $\P^{k-1}$ into two children. For example, if we consider an NDP on $[0,1]$ generated from sequantially dividing sets into two halves in the middle of the interval, then we have an NDP such that for $k\geq 0$,  $\P^{k}=\left\{\left[\frac{l-1}{2^k},\frac{l}{2^k}\right)\right\}_{l\in\{1,...,2^k\}}$. We refer to this particular NDP as the {\em canonical} NDP, and note that $\bigcup \P^{k}$ generates the Borel $\sigma$-algebra. In the following, we shall consider only NDPs that generate the Borel $\sigma$-algebra.
Now let us assume that each dimension of $\om$ has a corresponding NDP. For our purpose, the NDP for each dimension can be distinct, but for ease of illustration let us assume that they are all the canonical NDPs on $[0,1]$. We consider the cross-products of these marginal NDPs on each dimension, which creates a cascade of partitions on the joint sample space. Specifically, for any vector of non-negative integers $\bk=(k_1,...,k_D)\in\N_0^D$, $\P^{k_1}\times\cdots\times\P^{k_D}$ forms a partition of $\om$. The elements of this partition are rectangles of the form
\[
A=A_1\times A_2 \times \cdots \times A_D, \quad \text{with $A_d \in \P^{k_d}$ for all $d=1,2,\ldots, D$.}
\]
Note that the vector $\bk$ encodes the level of $A$ in the NDP for each dimension of $\om$. That is, $k_d$ is the level of the NDP on $[0,1]$ to which the $d$th margin of $A$ belongs. 
From now on, we shall refer to 
a set $A$ of the above form as a {\em cuboid}. 
We refer to the sum of all $k_d$, $r=\sum_{d=1}^{D} k_d$, as the {\em resolution} of $A$. 
{\bf Figure~\ref{fig:cuboid-strat}} illustrates 
a cuboid $A$ of resolution 3 in a 3-dimensional sample space with canonical NDPs on the margins.

\begin{figure}[!h]
\centering
\includegraphics[width=0.5\textwidth]{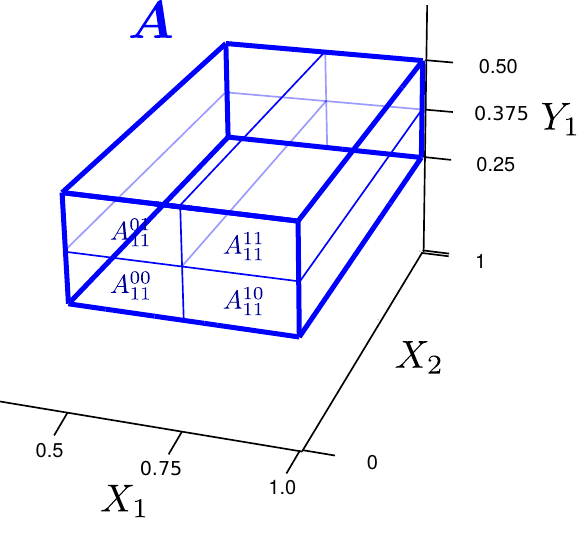}
\caption{\label{fig:cuboid-strat-Aij-ab} The division of the cuboid $A$ in Figure~\ref{fig:cuboid-strat} into four blocks along dimension~1 for $\bX$ and dimension~1 for $\bY$.
}
\end{figure}

We are now ready to construct the $2\times 2$ tables on which to carry out univariate tests of independence. 
One can divide a cuboid $A$ into four blocks according to the NDP along any pair of its margins 
while keeping all other dimensions intact. For the division involving dimension $i$ of $\bX$ and dimension $j$ of $\bY$, 
we use $A_{ij}^{00}$, $A_{ij}^{01}$, $A_{ij}^{10}$, and $A_{ij}^{11}$ to denote these four blocks.
{\bf Figure~\ref{fig:cuboid-strat-Aij-ab}} illustrates a division on the cuboid demonstrated in {\bf Figure~\ref{fig:cuboid-strat}}.

Suppose now that $F$ is the joint sampling distribution of $(\bX,\bY)$, then for the $2\times 2$ division of $A$ along the $i$th dimension of $\bX$ and $j$th dimension of $\bY$, we can define a corresponding odds-ratio that characterizes the dependency in $F$ on the $2\times 2$ division,
\[
\theta_{ij}(A) = \frac{F(A_{ij}^{10})F(A_{ij}^{01})}{F(A_{ij}^{00})F(A_{ij}^{11})}.
\]
An i.i.d.\ sample from $F$ will give rise to a $2\times 2$ contingency table formed by the number of data points lying in the four blocks
\[ \{n(A_{ij}^{00}), n(A_{ij}^{01}), n(A_{ij}^{10}), n(A_{ij}^{11})\}\quad \text{or} \quad
\begin{tabular}{|c|c|}\hline
$n(A_{ij}^{00})$ & $n(A_{ij}^{01})$\\\hline
$n(A_{ij}^{10})$ & $n(A_{ij}^{11})$\\\hline
\end{tabular}\]
where $n(A)$ represents the number of data points in $A$. 

\begin{figure}[!ht]
\centering
\begin{subfigure}{.33\textwidth}
\centering
	\includegraphics[width=1\textwidth]{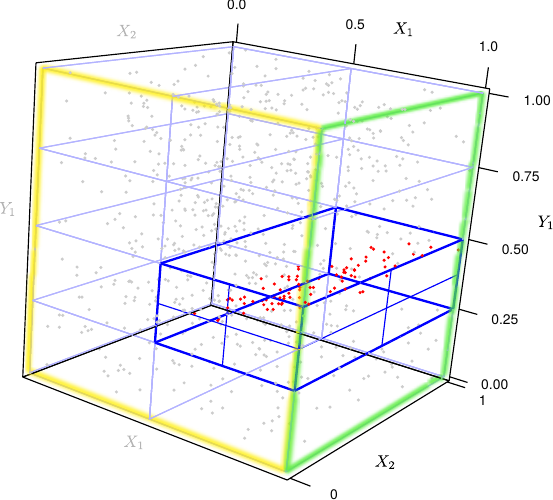}
	\caption{}
	\end{subfigure}
	\begin{subfigure}{0.66\textwidth}
	\centering
	\includegraphics[width=1\textwidth]{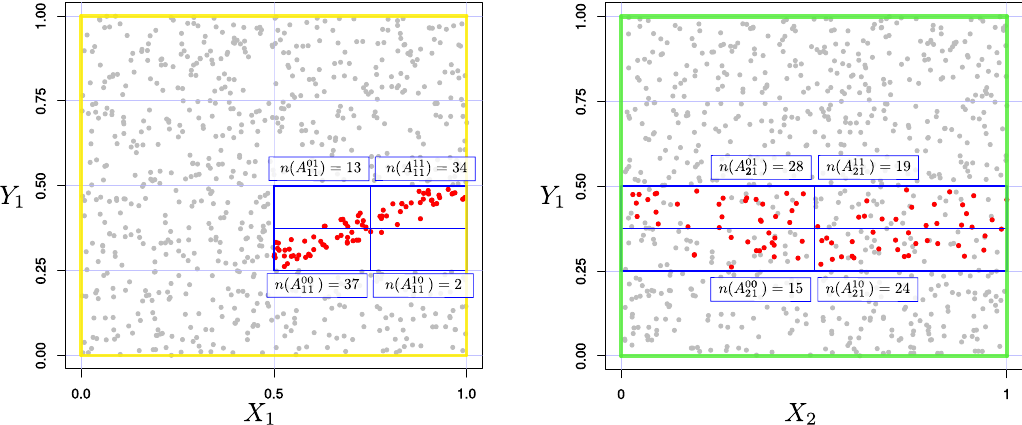}
	\caption{}
	\end{subfigure}

\caption{\label{fig:mvar-win-strat}Illustration of the two $2\times 2$ contingency tables on a cuboid $A$ arising from an i.i.d.\ sample in which dependency exists in $(X_1,Y_1)$}.
\end{figure}

One can test whether $\theta_{ij}(A)=1$ based on this contingency table. 
While several standard tests are available for testing independence on a $2\times 2$ table, we adopt Fisher's exact test. As we will show in Section~\ref{sec:theory}, it turns out that the conditional nature of Fisher's test plays a crucial role in our finite-sample theory---it ensures that the resulting testing procedure obtains exact validity at any finite sample size without resampling or asymptotics. {\bf Figure~\ref{fig:mvar-win-strat}} illustrates two contingency tables on which Fisher's test is applied for a cuboid $A$. In the following, we will use $p_{ij}(A)$ to represent the resulting $p$-value from the test on this particular $2\times 2$ table. 

How does testing those ``local'' nulls $\theta_{ij}(A)=1$ relate to our original ``global'' hypothesis of $\bX \indep \bY$? It is obvious that if $\bX\indep \bY$ then independence must hold---that is, $\theta_{ij}(A)=1$---for any $A$ and any pair of $X$-$Y$ margins $i$ and $j$. However, the reverse is not obvious---does independence on these $2\times 2$ tables formed under the marginal NDPs also imply that $\bX$ and $\bY$ are independent? If this is the case, then one can test for independence between $\bX$ and $\bY$ by testing whether $\theta_{ij}(A)=1$ on the $2\times 2$ tables. 
The next theorem confirms that this is indeed the case.
\begin{theorem}
\label{thm:thm0}
 $\bX\indep \bY$ 
if and only if 
$\theta_{ij}(A)=1$ for all pairs of dimension $i$ of $\bX$ and dimension $j$ of $\bY$ on all cuboids $A$.
\end{theorem}

This theorem implies that one can in principle test for independence between two random vectors $\bX$ and $\bY$ by exhaustively testing whether independence holds on each of the $2\times 2$ tables constructed on all cuboids up to some maximum resolution, aimed at identifying dependency structures up to a certain level of detail.  
This boils down to a multiple testing problem involving a collection of $p$-values computed on all of the $2\times 2$ tables up to the maximal resolution. However, such a brute-force exhaustive scan is not practical when the dimensionality grows. If one were to exhaustively test independence on all possible $2\times 2$ tables of all cuboids up to even just a moderate resolution, the number of tests required would quickly become prohibitive.
Specifically, the total number of tests to be completed up to a resolution of $R$ is $\sum_{\rho=0}^R  D_x\cdot D_y\cdot 2^\rho\cdot {{\rho+D-1}\choose{D-1}}$. 

For multivariate problems of more than a handful of dimensions then, one must be selective in carrying out the univariate tests.
Beyond the consideration of computational practicality, reducing the number of tests is also desirable for the sake of statistical performance. Every additional test comes with a price in multiple testing control, and thus it is important to be discreet in choosing the tests to complete.

\subsection{MultiFIT: a coarse-to-fine adaptive testing procedure}
\label{sec:algo}
\begin{figure}[ht!]
\centering
\includegraphics[width=0.75\textwidth]{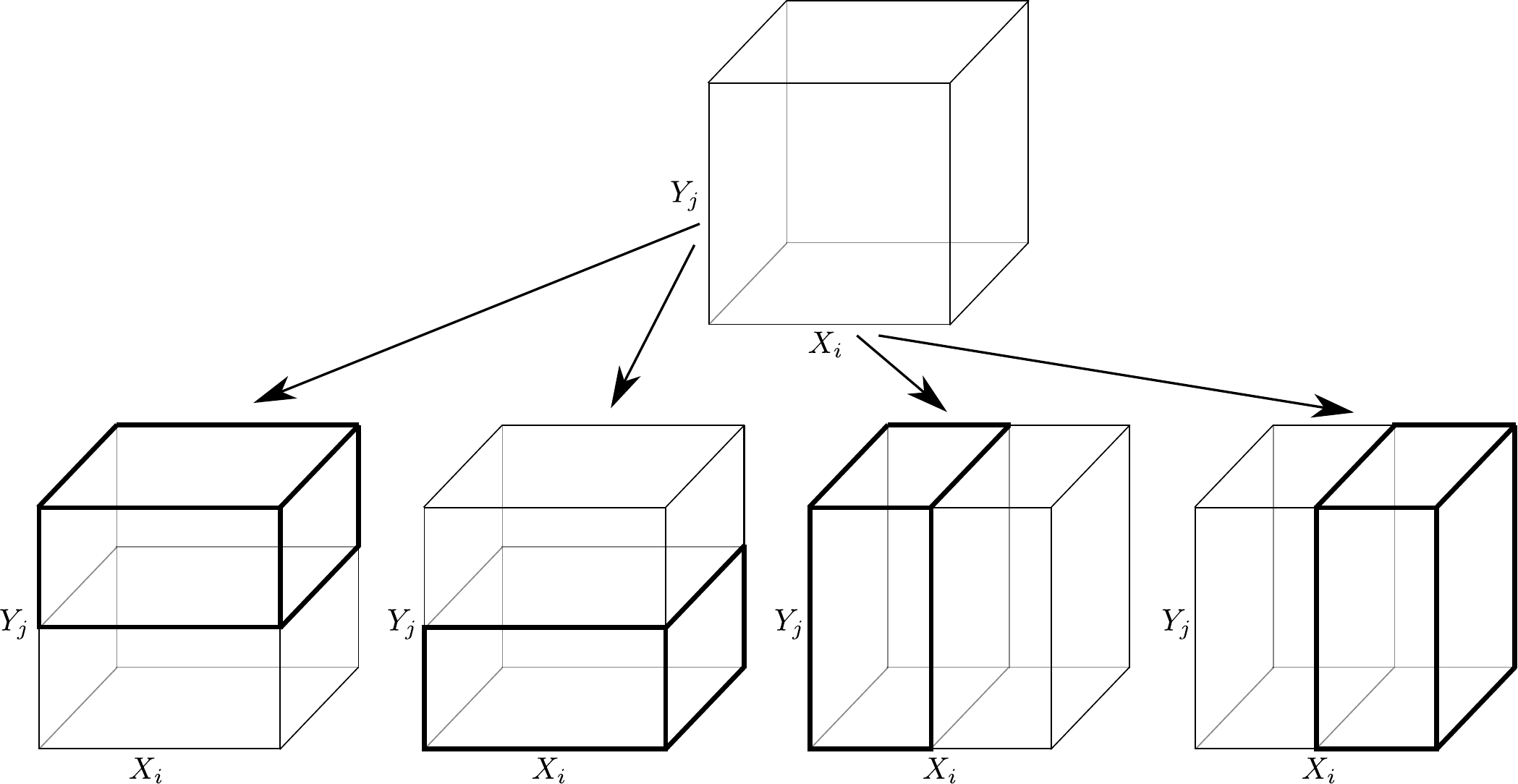}
\caption{The selection of tables for testing based on the statistical evidence on their parent. The two right children correspond to dividing $A$ along the margin that corresponds to the $i$th margin of $\bX$, and the two left children correspond to dividing $A$ along the margin that corresponds to the $j$th margin of $\bY$. Those four children are tested in resolution $r+1$ if their parent $A$ in resolution $A$ produces a $p$-value below the threshold $p^*$.\label{fig:parent-children}
}
\end{figure}

Given the above considerations, 
we propose a data-adaptive strategy that selects in each resolution a subset of the available tables to test based on the statistical evidence attained on coarser resolutions. In particular, only the ``children'' of tables in the previous resolution whose $p$-values are below a pre-specified threshold are selected for testing. {\bf Figure~\ref{fig:parent-children}} provides an illustration. Suppose that cuboid $A$ in resolution $r$ satisfies $p_{ij}(A)<p^{*}$, some preset threshold, then the four children cuboids, generated by dividing $A$ in the $i$th or the $j$th dimensions are tested in resolution $r+1$. 
This coarse-to-fine testing procedure terminates at a maximal resolution $R_{max}$ or when no cuboids at the current resolution have $p$-values passing the threshold.

The rationale behind this strategy is to exploit the spatial smoothness of dependency structures---when $\bX$ and $\bY$ are dependent, adjacent and nested cuboids tend to contain empirical evidence for the dependency in a ``correlated'' manner. (It is worth noting that here the ``correlation'' corresponds to our assumption about the underlying sampling distribution that its dependency structure is spatially smooth, not the sampling behavior of the data points given the sampling distribution.) Thus using the statistical evidence at coarser resolutions to inform which cuboids to test in finer resolutions can lead to effective detection of the dependency structure.

Next we formally present the adaptive testing procedure. We let $\C^{(r)}$ denote the collection of cuboids at resolution $r$ on which we carry out independence tests over all of the corresponding $D_x\cdot D_y$  $2\times 2$ tables, one for each $(i,j)$ pair of margins, where $i$ and $j$ are the indices for the $X$ and $Y$ margins respectively.
The procedure consists of three components:
\begin{enumerate}[1.]
\setcounter{enumi}{-1}

\item {\bf Initialization}: Let $\C^{(0)}$ be $\om$, and let $\C^{(r)}=\emptyset$ for $1\leq r\leq R_{max}$. 

\item {\bf Coarse-to-fine scanning}: For $r=0,1,2,\ldots,R_{max}$ do the following:
\begin{itemize}
\item[1a.] {\bf Independence testing}: Apply Fisher's exact test of independence to the $D_x\cdot D_y$ $2\times 2$ tables of each cuboid $A\in \C^{(r)}$ and record the $p$-values. 
\item[1b.] {\bf Selection of cuboids to test for the next resolution}: When $r < R_{max}$, 
if the $(i,j)$-table for a cuboid $A\in\mathcal{C}^{(r)}$ has a $p$-value more significant than a threshold $p^*$, add to $\C^{(r+1)}$ the four \textit{child cuboids} of $A$ generated from dividing $A$ along the $i$th and the $j$th dimensions respectively, each generating two children.
\end{itemize}
\item {\bf Multiple testing control}: Apply any valid multiple testing control procedure on the entire set of $p$-values generated by the algorithm, thereby controlling the level of the entire testing procedure at $\alpha$.
\end{enumerate}

Although the data-adaptive selection in {\bf Step~1b} is designed to overcome the explosive number of tests required when the dimensionality is large, it is still often feasible to apply exhaustive testing up to some resolution $R^*<R_{max}$. In other words, one can test on all available cuboids up to resolution~$R^*$, and let the adaptive selection of the cuboids in {\bf Step~1b} kick in for resolutions beyond $R^*$. 
In our software, we allow the user to specify a resolution $R^*$ below which exhaustive testing is adopted.  
A smaller value for $R^*$ will favor the detection of more global signals, while a larger $R^*$ will favor localized signals.

In our implementation of the testing procedure, we consider two different approaches for achieving the multiple testing control in {\bf Step~2}.

{\em Strategy I. A holistic approach to multiple testing.} Under this strategy, one applies multiple testing control procedure on the {\em entire set of $p$-values} generated in {\bf Step~1} of \texttt{MultiFIT} {\em all at once}, regardless of the resolution of the corresponding table. Simple choices of the multiple testing devices include Bonferroni are Holm corrections.

{\em Strategy II. A resolution-specific approach to multiple testing.} Under this strategy, one applies multiple testing control in two stages---first on the $p$-values within each resolution level, producing an intermediate, intra-resolution significance level for each resolution, and then in the second stage further correct these intra-resolution ``$p$-values'' over all the resolutions, which will produce a valid, corrected overall $p$-value for testing the global null hypothesis of independence. This strategy has the benefit that one can now ``allocate'' a fixed level budget to each resolution, and thus avoids the possibility of loss in power due to having many more tables tested in high resolutions than coarse ones. This method is generally more powerful than the holistic approach above for testing the global null hypothesis when a dependency structure exists in coarser resolutions.

An additional benefit of the resolution-specific approach is that it can be implemented with early stopping so that the \texttt{MultiFIT} procedure
can terminate as soon as there is sufficient evidence for rejecting the global null in the first few resolutions without continuing into testing on higher resolutions. This is possible because in this approach we bound the influence of tables in finer resolutions on the (corrected) significance level of tests in coarser resolutions. Our software implements this early stopping strategy for the resolution-specific approach to multiple testing when Holm's method is used for intra-resolution correction along with Bonferroni's method for cross-resolution correction. Early stopping can reduce the time complexity significantly in the presence of a  global signal (see {\bf Figure~\ref{fig:scaling}}).

Detailed pseudo-code for the procedure is provided in Supplement~\ref{sec:algo_supp}. We call this testing procedure {\tt MultiFIT}, which stands for Multi-scale Fisher's Independence Test.

\subsection{Finite-sample validity and large-sample consistency}
\label{sec:theory}
Because \texttt{MultiFIT} formulates the test of independence as a multiple testing problem, its inferential validity rests on whether the $p$-values are indeed valid, i.e., that they are stochastically larger than a uniform random variable under the null hypothesis. Note that the $p$-values for the cuboids selected in the \texttt{MultiFIT} procedure are computed according to the (central) hypergeometric null distribution on the $2\times 2$ tables. At first glance, these null distributions appear to ignore the data-adaptive selection of a cuboid $A$ based on the evidence in its ancestral cuboids. As such, one may suspect that there might be a selection bias that causes such $p$-values to lose their face values. 

The following theorem and corollary resolve this concern by showing that, interestingly, the distribution of 
all the selected $2\times 2$ tables given their marginal totals are independent of the event that they 
are selected in the procedure, and hence the $p$-values computed in the procedure are indeed still valid despite the adaptive sequential selection. Consequently, one can indeed control the level of the entire procedure using multiple testing methods based on these $p$-values.

\begin{theorem}
\label{lem:lem3}
Under the null hypothesis $\bX\indep \bY$,
\[
n(A^{00}_{ij}) \indep \mathbf{1}(A\in \C^{(r)})\,|\, n(A^{0\cdot}_{ij}),n(A^{\cdot 0}_{ij}), n(A) 
\]
for all cuboids $A$ of resolution $r$ and all pairs $(i,j)$ of the margins, where $n(A^{0\cdot}_{ij})=n(A^{00}_{ij})+n(A^{01}_{ij})$ and $n(A^{\cdot 0}_{ij})=n(A^{00}_{ij})+n(A^{10}_{ij})$, and $\mathbf{1}(A\in \C^{(r)})$ is the indicator for the event that $A$ is selected to be tested in the {\rm \texttt{MultiFIT}} procedure.
\end{theorem}
In other words, for any cuboid $A$, the conditional distribution of the $2\times 2$ table on each pair of $X$-$Y$ margins given the corresponding marginal totals is the same central hypergeometric distribution when $\bX\indep \bY$ {\em whether or not} we condition on the event that cuboid $A$ is selected to be tested in the {\rm \texttt{MultiFIT}} procedure. As such, the $p$-values from the Fisher's exact tests applied on the adaptively selected tables in our procedure can be treated at face value, which justifies using multiple testing adjustment based on these $p$-values to control the level.

\begin{corollary}
\label{cor:cor1}
The $p$-values computed during {\bf Step~1} of the {\rm \texttt{MultiFIT}} procedure are valid, and thus {\bf Step~2} of the procedure can control the level of the entire testing procedure at any given level $\alpha$.
\end{corollary}

The above theorem and corollary provide a strong theoretical guarantee---unavailable to other existing methods---that \texttt{MultiFIT} attains exact control of the level at any finite sample size. This is an extremely important property in that for multivariate sample spaces traditional large $n$ asymptotic controls of the level can often be inaccurate, and existing methods typically appeal to resampling strategies such as permutation to provide approximate finite-sample control of the level. But permutation is often computationally prohibitive in this context in that even just a single run of a test can be expensive, not to mention applying the same test hundreds to thousands of times. In contrast, \texttt{MultiFIT} achieves exact control of the level by a single run of the procedure without resampling. We offer a numerical validation of level control through simulations in Section~\ref{sec:fwer}.

We also note that the proof of Theorem~\ref{lem:lem3}
turns out to be conceptually interesting and elucidates why the adoption of the Fisher's exact test on each $2\times 2$ table is critical to ensuring the exact finite sample validity of the {\tt MultiFIT} procedure. In particular, the event that a cuboid $A$ is selected to be tested in {\tt MultiFIT} is in the $\sigma$-algebra generated by the $p$-values on all of its ancestral cuboids, which can be shown to be independent of the counts in the $2\times 2$ table on $A$ under the null hypothesis of independence once the corresponding marginal totals are conditioned upon. This independence is elucidated under a Bayesian network representation of the multivariate central hypergeometric (CHG) distribution \cite[Theorem~3]{mamao2017}. Accordingly, conditioning on the selection of a cuboid under {\tt MultiFIT} does not alter the null distribution of the $p$-values for the $2\times 2$ tables on that cuboid, and thus the validity of the procedure is maintained even with the adaptive selection of the tables to test on. Below we provide a sketch of the proof for Theorem~\ref{lem:lem3} for interested readers and defer the technical details to Supplement~\ref{sec:supplement-proofs}.

{\em Sketch of Proof for Theorem~\ref{lem:lem3}}: 
For two non-negative integers $a$ and $b$, let $\mathbf{n}_{a, b}$ denote a $2^{a}\times2^{b}$ contingency table formed by a cross-product of a marginal partition on $\bX$ at depth $a$ and a marginal partition on $\bY$ at depth $b$. Specifically, it is the $2^a\times 2^b$ contingencey table corresponding to a partition
$\P^{k_1}\times \cdots \times \P^{k_D}$ of $\om$,
where $\sum_{d=1}^{D_{x}} k_d = a$ and $\sum_{d=D_{x}+1}^{D} k_d = b$. Under the null hypothesis that $\bX\indep \bY$, the sampling distribution of any such table $\mathbf{n}_{a,b}$ {\rm given all of its row totals and column totals} is a multivariate CHG distribution.

By Theorem~3 in \cite{mamao2017}, 
a draw from the central multivariate hypergeometric distribution such as  $\mathbf{n}_{a,b}$ can actually be generated inductively from coarse-to-fine resolutions using univariate CHG distributions. Specifically, suppose we have already generated 
the table $\mathbf{n}_{a-1,b}$ and $\mathbf{n}_{a,b-1}$, then the conditional distribution of $\mathbf{n}_{a,b}$ given its row and column totals, as well as the two ``parent'' tables $\mathbf{n}_{a-1,b}$ and $\mathbf{n}_{a,b-1}$, are simply a collection of independent univariate central hypergeometric distributions---one for each adjacent $2\times 2$ subtables in $\mathbf{n}_{a,b}$ given its row totals and column totals, which correspond to cell counts in $\mathbf{n}_{a-1,b}$ and $\mathbf{n}_{a,b-1}$.

\begin{figure}[t]
\centering
    \includegraphics[width=0.95\textwidth]{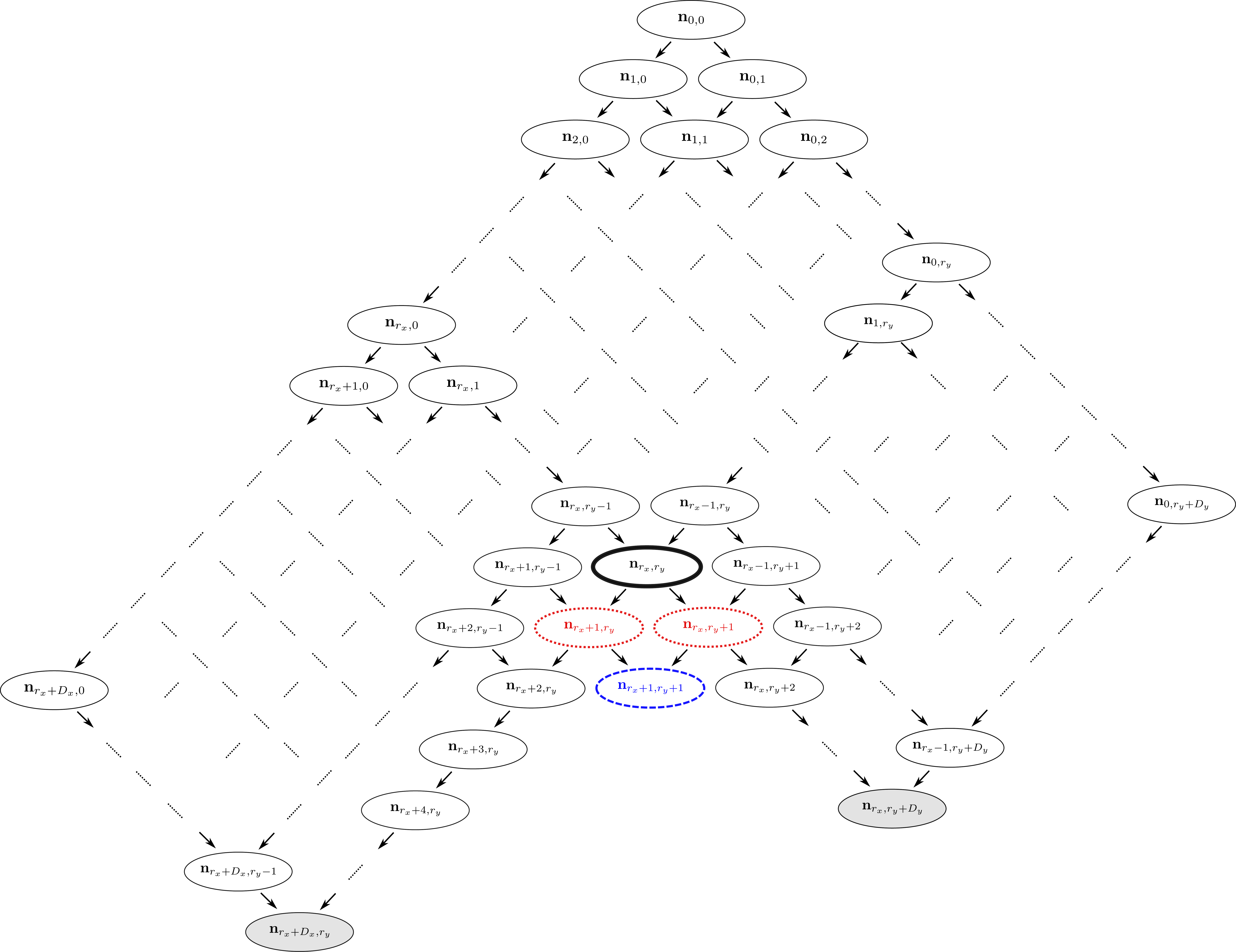}
\caption{
\label{fig:graph}
A Bayesian network augmentation for the multivariate CHG model on contingency tables formed by cross-products of sequential marginal partitions on $\om_{\bX}$ and those on $\om_{\bY}$.}
\end{figure}

Let $A$ be a cuboid that arises from dividing the $\bX$ margins a total of $r_x$ times and the $\bY$ margins a total of $r_y$ times.
The above reasoning implies that one can show by construction that for any $2\times 2$ table on a cuboid $A$, there exists a Bayesian network in the form presented in {\bf Figure~\ref{fig:graph}} such that the total number of observations in $A$, $n(A)$, is an element in the contingency table $\bn_{r_x,r_y}$ (the node with bold black boundary in Figure~\ref{fig:graph}), the counts for the four blocks of the $2\times 2$ table, $n(A_{ij}^{00})$, $n(A_{ij}^{01})$, $n(A_{ij}^{10})$, and $n(A_{ij}^{11})$, are in $\bn_{r_x+1,r_y+1}$ (the node with blue dashed boundary in Figure~\ref{fig:graph}), and the marginal totals of $A$ are in $\bn_{r_x+1,r_y}$ and $\bn_{r_x,r_y+1}$ (the two nodes with dotted red boundaries in Figure~\ref{fig:graph}). In addition, the counts of all of the $2\times 2$ tables on ancestors of $A$ are measurable with respect to the $\sigma$-algebra generated the gray-shaded nodes in the Bayesian network, and thus are independent of the $2\times 2$ table on $A$ given the marginal totals.
Therefore the selection of a table does not influence the null distribution once the marginal totals are conditioned upon, as such conditioning blocks all the paths from these ancestral nodes to the blue dashed node.

\vspace{1em}

Now that we have established the finite-sample exact validity of the \texttt{MultiFIT} procedure, our last theoretical result shows that when the sample size $n$ grows, 
under certain conditions \texttt{MultiFIT} can consistently reject the null hypothesis of independence.

\begin{theorem}[Large-sample consistency]
\label{thm:local_consistency}
Suppose $\bX$ and $\bY$ are not independent under their sampling distribution $F$. Let $(\bX_1,\bY_1),(\bX_2,\bY_2),\ldots,(\bX_n,\bY_n)$ be i.i.d.\ observations from $F$. As $n\rightarrow \infty$, 
suppose one of the following is true
\begin{itemize}
\item[(i)] $R^*$ is fixed but large enough such that there exists at least one cuboid $A$ of resolution $r\leq R^*$ with $\theta_{ij}(A)\neq 1$ for some pairs of margin $(i,j)$,
\item[(ii)] $R^* \rightarrow \infty$ and it is $o(\log n)$.
\end{itemize}
Then the power for {\tt MultiFIT} to reject the null hypothesis that $\bX \indep \bY$ converges to~1. 
\end{theorem}

\subsection{Practical considerations in applying {\tt MultiFIT}}
We close this section by discussing some practical aspects in applying the {\tt MultiFIT} procedure. 
We set the default value for the p-value threshold $p^*$ in our software for resolutions higher than $R^*$ at $(D_x\cdot D_y\cdot \log_2(n))^{-1}$. This keeps the number of $2\times 2$ tables tested constant (on average) under the null hypothesis irrespective of the number of dimensions, while also making the threshold more stringent with increasing sample size in such a way that makes the total number of tables scales roughly linearly with the sample size, which we will confirm numerically in the next section.

We note that under this strategy of setting $p^*$, 
we get that for certain alternatives, in particular those that are pervasive over the sample space and involve a large number of cuboids, the complexity of the \texttt{MultiFIT} procedure may be higher than $O(n\log n)$. Such large-scale, global alternatives, however, can usually be detected in coarse resolutions, and thus in practice when the algorithm is equipped with early stopping it will in fact run faster with larger $n$ under such alternatives. 
If the practitioner wishes to ensure a strict $O(n\log n)$ bound on the computational complexity with or without incorporating early stopping, 
a simple approximate version of the \texttt{MultiFIT} algorithm can achieve this.  Specifically, in {\bf Step~1b} of \texttt{MultiFIT}, instead of including child cuboids of {\em all} cuboids $A$ with p-value less than $p^*$, we can include only child cuboids with p-value less than $p^*$ up to a maximum number of cuboids $A$ (e.g., 100) with the smallest Fisher's p-values. This alternative constraint ensures that the computational cost of \texttt{MultiFIT} algorithm is strictly bounded at $O(n\log n)$. 
While under this approximation the conditions for ensuring the finite-sample guarantees are no longer satisfied, we found in practice that its statistical power ({\bf Figure~\ref{fig:power}} and {\bf Figure~\ref{fig:power_spread}}) and level ({\bf Figure~\ref{fig:fwer}}) hardly differs from those of the exact \texttt{MultiFIT} procedure in essentially all of the numerical settings we have encountered.

\section{Numerical Examples}
\label{sec:numerical}

\subsection{Computational Scalability}\label{subsec:scaling}
Because computational scalability is a key motivation for our approach, we start by evaluating the computational scalability of {\tt MultiFIT} with those of three other state-of-the-art methods with well-documented software---the Heller-Heller-Gorfine (\texttt{HHG}) multivariate test of association from \cite{heller2013}, the Distance Covariance (\texttt{DCov}) method of \cite{szekely2009}, and the kernel-based method (\texttt{dHSIC}) of \cite{pfister2018}.

We apply these methods to data sets simulated under six scenarios described in {\bf Table~\ref{tbl:scenarios}} along with a ``null'' scenario where there is no dependence. Here we report the results for two scenarios as they represent the best- and worst-case computational scenarios for {\tt MultiFIT} and defer the rest of the scenarios to {\bf Figure~\ref{fig:scaling_supp}} in the Supplementary Materials. 
The first scenario we report involves data generated under the null hypothesis, with all margins being drawn independently from a standard normal distribution. Under the second scenario, one dimension of $\bY$ is strongly correlated with a dimension of $\bX$ under the ``linear'' scenario from {\bf Table~\ref{tbl:scenarios}} with $l=3$. While in practice non-linear alternatives are the main motivation for the nonparametric tests being considered here, the linear scenario is essentially the worst-case scenario for {\tt MultiFIT} in terms of computational time. 
The reason is that the stronger the dependency at coarser levels, the more tests will be performed under {\tt MultiFIT} because more tests will pass the $p$-value threshold at coarser levels. As such, these two scenarios represent the two ends of the spectrum in the amount of computation incurred under {\tt MultiFIT}. 

\begin{figure}[!ht]
    \centering
       \includegraphics[width=0.9\textwidth]{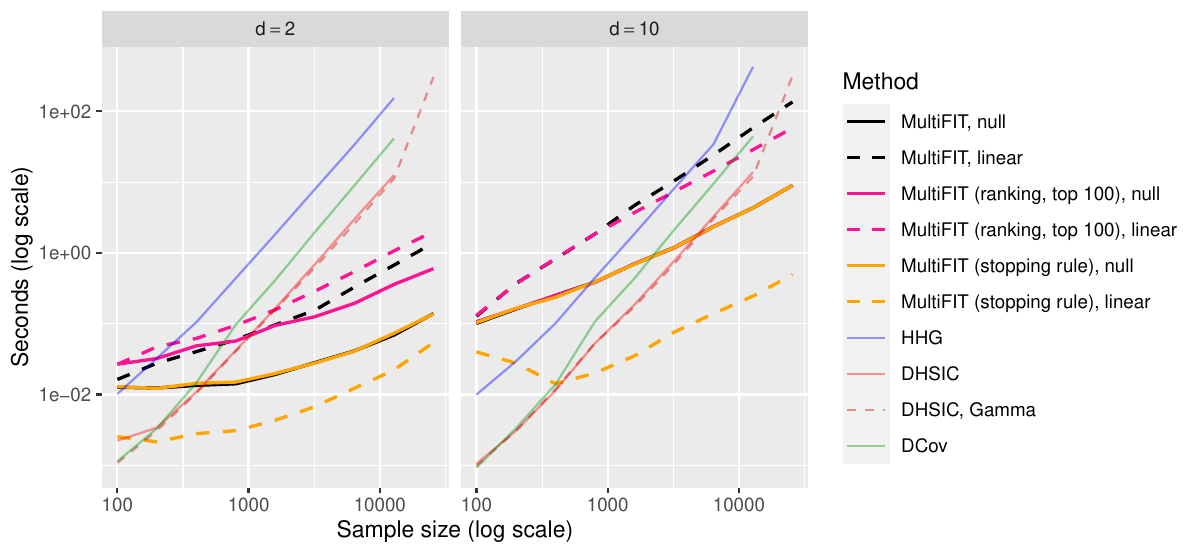}
  \caption{Computational scalability: a comparison of \texttt{HHG}, \texttt{DCov}, \texttt{dHSIC} (a single computation of the test statistic) and \texttt{MultiFIT} with $D_x=D_y=d$, 
  log runtime versus log sample size. \texttt{MultiFIT} was run with $R^*=1$ and $p^* = (D_x\cdot D_y\cdot \log_2(n))^{-1}$. Three variants of \texttt{MultiFIT} are investigated: the full algorithm, the approximate algorithm that keeps up to 100 most significant $p$-values at each resolution, and the full algorithm with early stopping.
  \texttt{MultiFIT} and {\tt dHSIC} with Gamma approximation does not require permutation. The other methods require permutations for level control and the reported time is for a single permutation}. \label{fig:scaling} 
\end{figure}

\textbf{Figure~\ref{fig:scaling}} plots the computational time versus the sample size (
in log-log scale) at different dimensionalities---2 and 10. All methods were run on the same desktop computer with a single Intel\textsuperscript{\textregistered} Core(TM) i7-3770 CPU unit at 3.40GHz, and the three competitors were evaluated up to the maximum sample size allowed by the available 16G RAM. We present the average duration of 10 executions of each method under different dimensions, $d=2$ and $d=10$. It is worth noting that the results for 
the competitors are for only a single permutation
while at least hundreds of resampling repetitions are required in order to perform inference.

Overall, the computational advantage \texttt{MultiFIT} is substantial---it scales approximately $O(n\log n)$ in sample size, while \texttt{HHG}, \texttt{DCov} and \texttt{dHSIC} without the Gamma approximation scale approximately $O(n^2)$. The Gamma approximation method of \texttt{dHSIC} makes the method faster in the presence of a strong signal, but it still cannot handle the larger sample sizes due to its memory requirement. 
\texttt{MultiFIT} with early stopping achieved the best computational efficiency at moderate to large sample sizes uniformly across non-null scenarios. As expected, early stopping does not reduce computation under the null. The approximate \texttt{MultiFIT} with a maximum number of cuboids per resolution on the other hand bounds the complexity by $O(n\log n)$.

We do acknowldge that the three competitors scale linearly in dimensionality while \texttt{MultiFIT} scales quadratically with the number of dimensions. As such \texttt{MultiFIT} is not suited for very high-dimensional problems. It is most suitable for problems up to tens of dimensions with large sample size.

\subsection{Power Comparison}
\label{sec:power}
We next examine the statistical power of the competing methods under several representative dependency scenarios. We consider two sets of simulation settings. In one set, the dependency exists only in a small number of margins, and thus is amenable to \texttt{MultiFIT}'s search over pairs of axes-aligned boundaries. In the other set, the dependency is spread over a large number of dimensions and thus is particularly adversarial to \texttt{MultiFIT}.

\begin{figure}[p]
    \centering
      \includegraphics[width=0.95\textwidth]{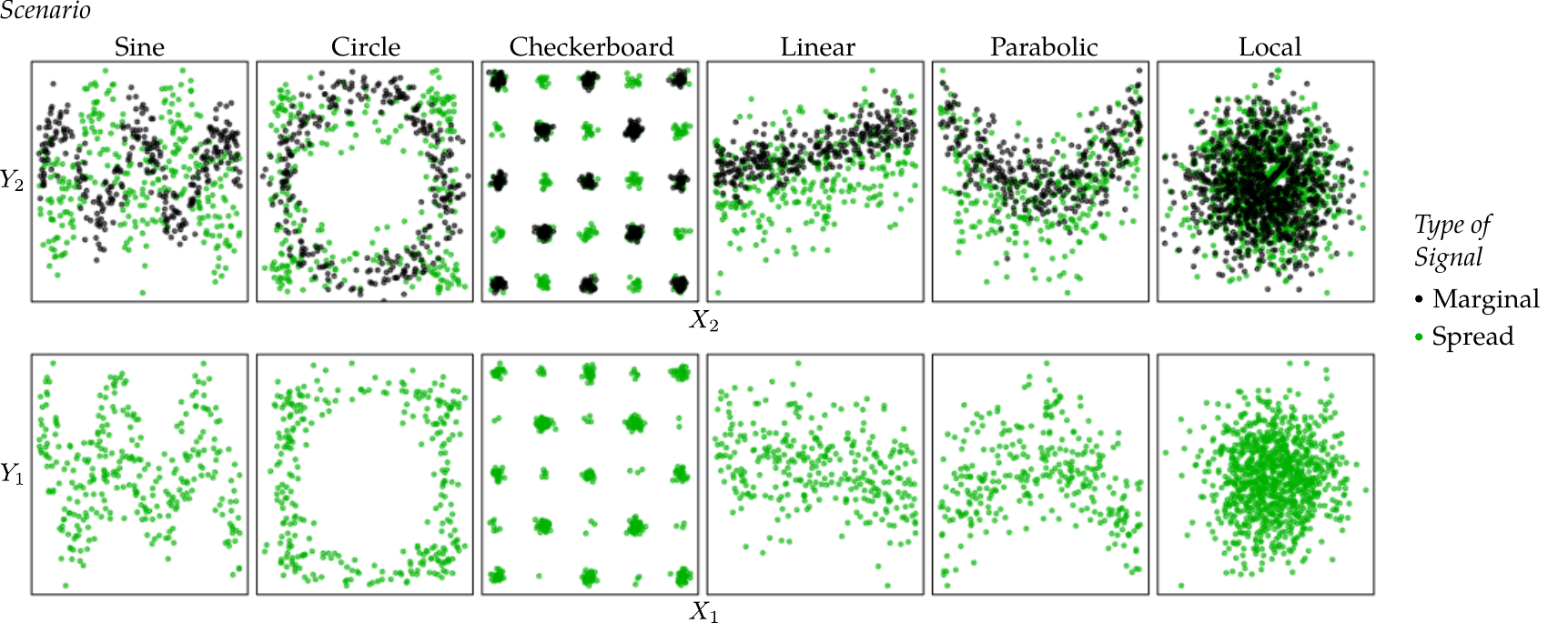}
      \caption{Visualization of the dependent margins of six scenarios with noise level 2. Note that the black, ``marginal'' scenario is only plotted in the top row as its $X_1$-$Y_1$ margins do not involve an interesting dependency, whereas the ``spread'' scenario is plotted in both. The dependency in the marginal scenario is more noticeable in the $X_2$-$Y_2$ margins than the spread scenario.}\label{fig:scenarios}
\end{figure}

In the first set of simulations, we let $X_1$ and $Y_1$ be independently normally distributed, whereas $X_2$ and $Y_2$ are dependent according to several different scenarios, which are illustrated in \textbf{Figure~\ref{fig:scenarios}} in black points in the upper row of plots and detailed in \textbf{Table~\ref{tbl:scenarios}}.  {\tt MultiFIT} has a natural advantage to detect such marginal dependencies as it focuses on the testing of pairs of margins.

In the second set of dependency scenarios, the true signal embodies dependencies of the $Y$ margins on multiple $X$ margins in terms of linear combinations or mixtures. This dissipates the strength of the dependency over many pairs of margins and thus is highly unfavorable to {\tt MultiFIT}. This set of scenarios is illustrated in \textbf{Figure~\ref{fig:scenarios}} in green points in the two rows of plots and detailed in \textbf{Table \ref{tbl:scenarios_spread}} and \textbf{Table \ref{tbl:scenarios_spread2}}.

For all scenarios except the ``local'' scenarios, we set the level of resolutions up to which exhaustive testing is done, $R^*=2$, and for the ``local'' scenarios, where a signal is embedded in a small portion of the sample space, we set $R^*=4$ to ensure exhaustive coverage up to resolution 4. 
In Section~\ref{sec:sensitivity} we present a detailed sensitivity analysis on the effects of the tuning parameters $p^*$ and $R^*$ on the power of the test under the simulation settings.

We performed 500 simulations for each scenario and at 20 different noise levels, and applied the four methods at the 5\% level. We first applied a rank transform to each of the $D$ margins for the simulated data as this is the default under {\tt MultiFIT} and the competitors \texttt{HHG}, \texttt{DCov} and \texttt{dHSIC} also performed much better with the marginal rank transform.

\begin{figure}
	  \includegraphics[width=1\textwidth]{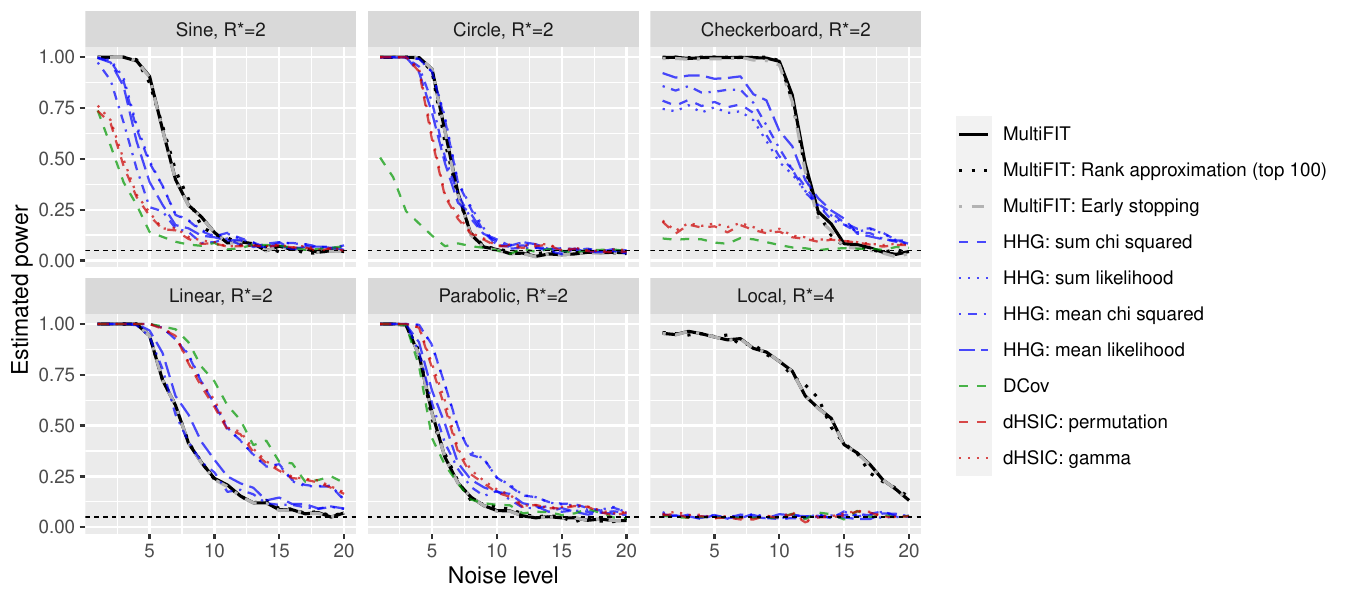}
	  \caption{Power versus noise level for different methods. Estimated power at 20 noise levels for the different methods under the six scenarios from {\bf Table~\ref{tbl:scenarios}}.\label{fig:power}}
\end{figure}

\begin{figure}
	  \includegraphics[width=1\textwidth]{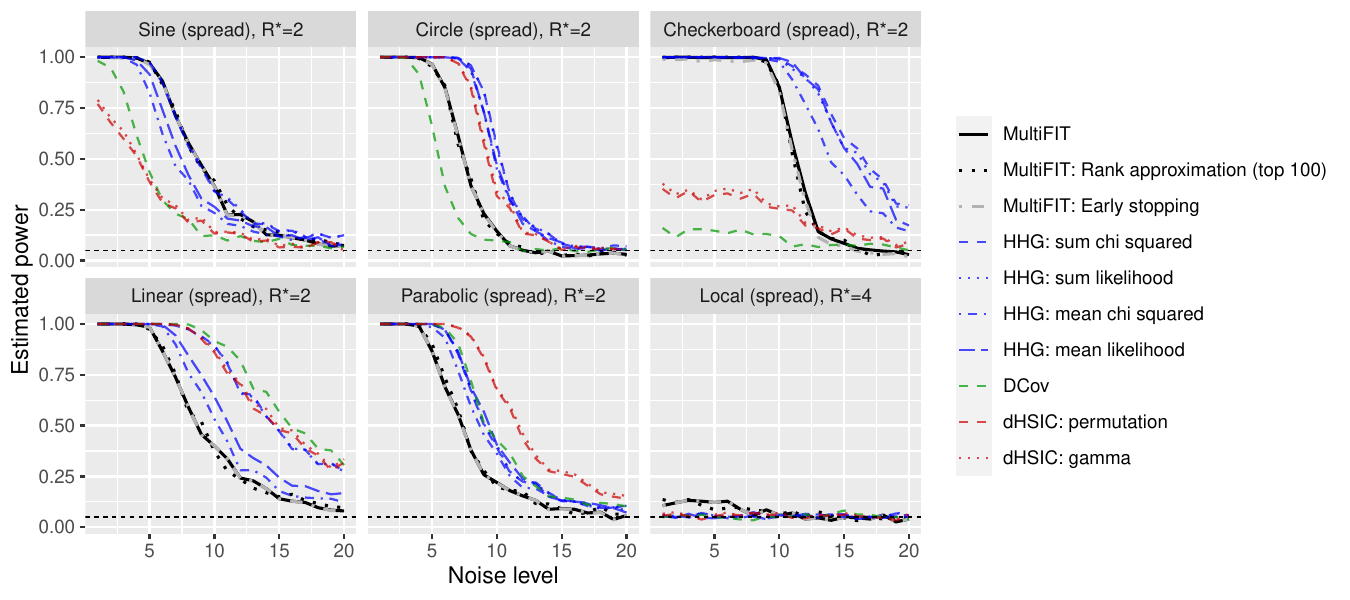}
	  \caption{Power versus noise level for different methods. Estimated power at 20 noise levels for the different methods under the six scenarios from {\bf Table~\ref{tbl:scenarios_spread}} and {\bf Table~\ref{tbl:scenarios_spread2}}.\label{fig:power_spread}}
\end{figure}

{\bf Figure~\ref{fig:power}} reports the result for the first set of simulations.  \texttt{MultiFIT} outperforms \texttt{HHG}, \texttt{DCov} and \texttt{dHSIC} for the ``sine'', ``circle'', ``checkerboard'' and ``local'' scenarios, the cases that are richer with local structures. For the more ``global'' dependency structures---``linear'' and ``parabolic''--- \texttt{HHG} and \texttt{dHSIC} outperform {\tt MultiFIT}, while \texttt{DCov} does so only in the ``linear'' case. This is explained by the fact that the signal is observable almost entirely in the coarsest level, and as we go into higher resolutions we merely add insignificant tests that reduce the overall power. 
In the second set of simulations ({\bf Figure~\ref{fig:power_spread}}), as expected, \texttt{MultiFIT} loses some power relative to the competitors. Nevertheless, its overall performance is still robust and it still outperforms all other methods in the ``sine'' and ``local'' spread scenarios.

The results are largely consistent with our intuition. Due to its divide-and-conquer nature, MultiFIT is particularly good a identifying dependency structures that concentrates within a small number of cuboids (i.e., local features), while its power is weaker when the dependency structure is spread over a large number of cuboids (i.e., global structures).

Finally, we acknowledg that the performance of some of the competitors, such as \texttt{HSIC}, could be further improved with more expert selection of tuning parameters. For example, the incorporation of a multi-scale bandwidth into \texttt{HSIC} \citep{li2019optimality} could further improve its performance.

\subsection{Learning the nature of the dependency}
\label{sec:rotated_circle}
So far we have focused on applying {\tt MultiFIT} for testing the null hypothesis of independence. In practice, especially in multivariate settings, the practitioner is often interested in not just testing the existence of dependence but to have an understanding of its nature. A by-product of the divide-and-conquer approach is the ability to shed light on the underlying dependency structure. In this section we provide two examples that illustrate \texttt{MultiFIT}'s ability of learning the nature of the dependency. In the first example we consider a dependency structure resulting from higher order interactions. In the second example the dependency consists of two sine waves in the $(X_1,Y_1)$ margin with different frequencies, while a third margin, $X_2$, determines the frequency. In both examples it is difficult to visualize the dependency in low-dimensional marginal visualizations. We show that after identifying the $2\times 2$ tables that contained statistically significant evidence for dependency (after multiple testing correction), by plotting the data points in those significant tables, one can learn and visualize the underlying dependency. In both examples, we use the holistic approach to multiple testing and adopt Holm's correction on the $p$-values.

\subsubsection{Example 1: Rotated 3D Circle}
\begin{figure}[!h]
\centering
\begin{subfigure}{.4\textwidth}
  \centering
  \frame{
  \includegraphics[width=0.9\textwidth]{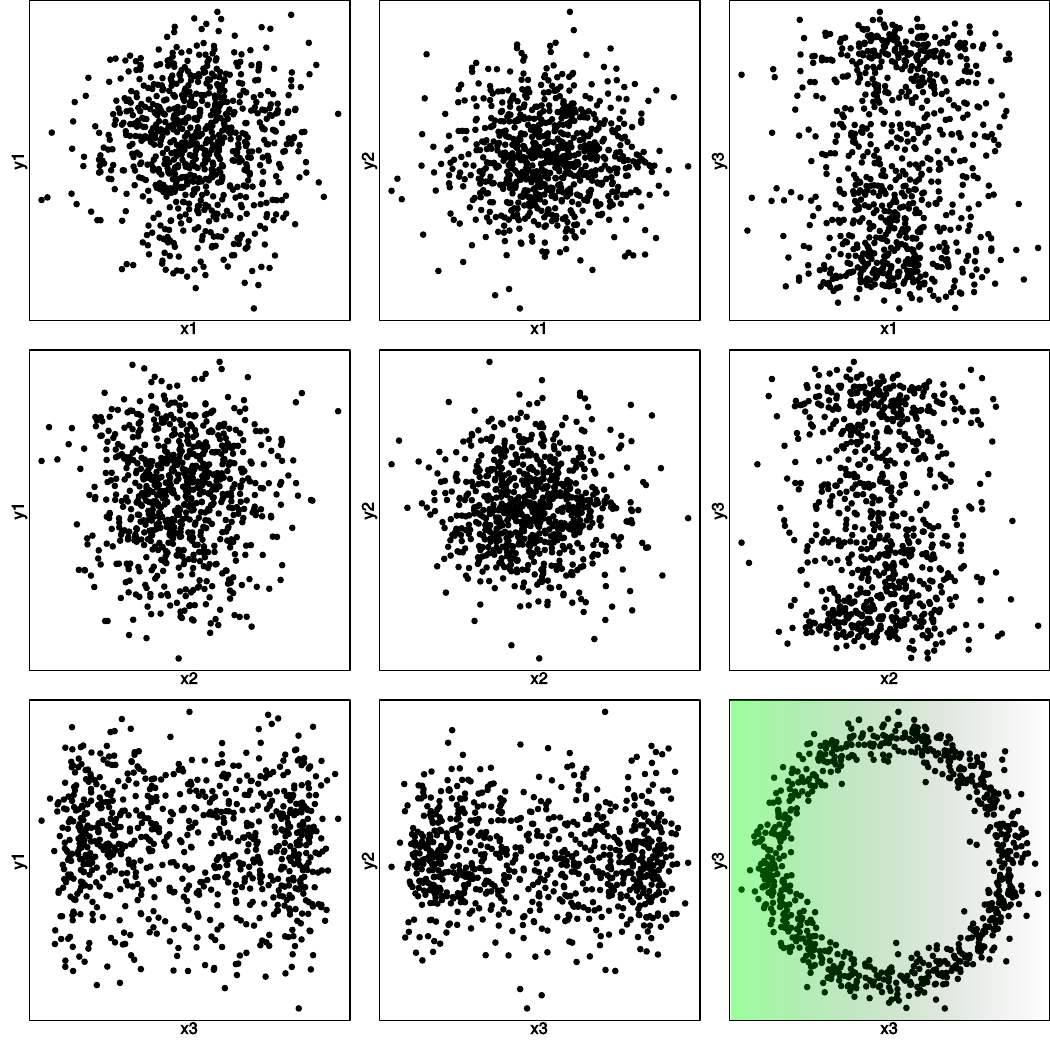}
  }
  \caption{}
  \label{fig:ex_circ_1}
\end{subfigure}%
\begin{subfigure}{.4\textwidth}
  \centering
  \frame{
  \includegraphics[width=0.9\textwidth]{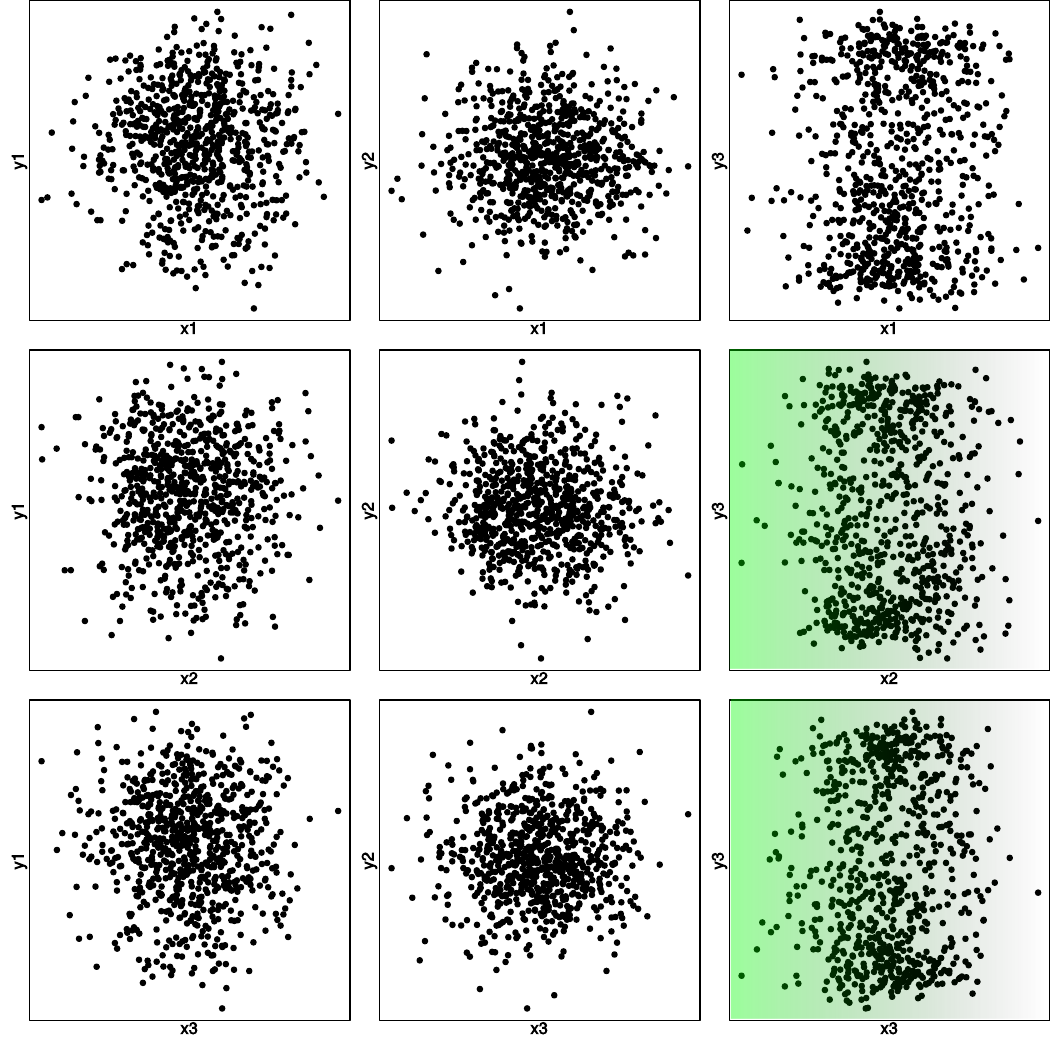}
  }
  \caption{}
  \label{fig:ex_circ_2}
\end{subfigure}
\caption{Marginal views of the data sample in Section~\ref{sec:rotated_circle} (a) before and (b) after rotation.\label{fig:exmplcircle} The dependency is easily visible in the marginal plots before rotation. Once rotated, the signal is spread among the margins and no longer visually obvious.}
\end{figure}

\begin{figure}[!h]
    \centering
	\includegraphics[width=0.8\textwidth]{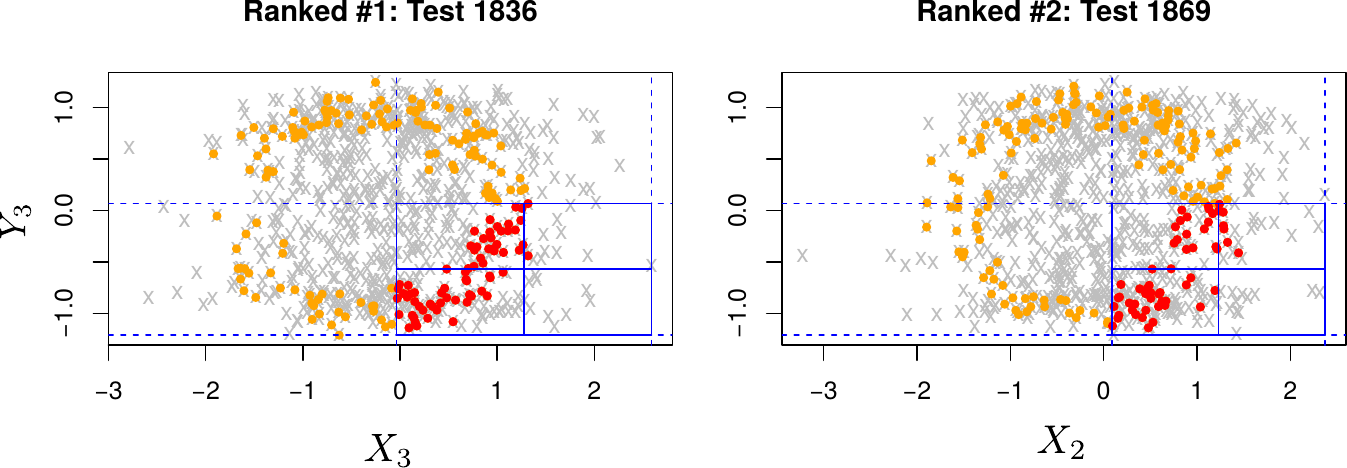}
\caption{Scatter plots for the observations in the three $2\times 2$ tables identified as most significant by {\tt MultiFIT} for the rotated circle scenario. (Significant tables are those with Holm's adjusted $p$-values below 0.001.) The dependency structure is again visible in the marginal views: red points are observations that are within  the cuboid that is tested, orange points are observations that are in a cuboid formed by expanding the tested cuboid so that the plotted margins are not subsetted. Notice how the left plot captures the dependency in the $X_3$-$Y_3$ plane while the right plot captures the dependency in the $X_2$-$Y_3$ plane. \label{fig:exmplsigcircle}}
\end{figure}

Let $\bX$ and $\bY$ each be of three dimensions, and simulate a sample with 800 observations. We first generate a ``circle'' scenario so that $X_1$, $Y_1$, $X_2$, and $Y_2$ are all i.i.d.\ standard normals, whereas $X_3=\cos(\theta)+\epsilon$, $Y_3=\sin(\theta)+\epsilon'$ where $\epsilon$ and $\epsilon'$ are i.i.d $\mathrm{N}(0,(1/10)^2)$ and $\theta\sim \mathrm{Uniform}(-\pi,\pi)$. We then rotate the circle by $\pi/4$ degrees in the $X_2$-$X_3$-$Y_3$ space by applying:
\begin{align*}
\left[\begin{matrix}
\cos(\pi/4) & -\sin(\pi/4) & 0\\
\sin(\pi/4) & \cos(\pi/4) & 0\\
0 & 0 & 1
\end{matrix}\right]\left[
\begin{matrix}
| & | & |\\
X_2 & X_3 & Y_3\\
| & | & |
\end{matrix}
\right].
\end{align*}
The rotated circle is no longer visible by examining the 2-dimensional margins. See \textbf{Figure~\ref{fig:exmplcircle}} for the marginal views of the sample before and after the rotation. \textbf{Figure~\ref{fig:exmplsigcircle}} plots the data points that lie in the $2\times 2$ tables identified as statistical significant (at 0.001 level after multiple testing adjustment with modified Holm's procedure) under the rotated setting. The underlying dependency pattern is clearly visible after selecting these tables. We found that in visualizing the identified tables, it is often useful to plot the data points that lie in the same slice of that table but with the full ranges of the plotted margins, as the identified table often captures a portion of the interesting dependency. \textbf{Figure~\ref{fig:exmplsigcircle}} demonstrates this technique by plotting those additional observations (in orange). For this reason, we have incorporated this plotting feature in our software.


\subsubsection{Example 2: Mixed Sine Signals}

Here we examine \texttt{MultiFIT}'s ability to detect a dependency structure consisting of two sine waves in different frequencies. 
Let $\bX=(X_1,X_2)'$ be a two-dimensional random vector 
with independent margins $X_1\sim U(0,1)$ and $X_2\sim Beta(0.3,0.3)$, and let

\[Y = \begin{cases}
\sin(10\cdot X_1) + \epsilon, & \text{if }X_2 > 0.5\\
\sin(40\cdot  X_1) + \epsilon, & \text{if }X_2\leq0.5
\end{cases}\]

\begin{figure}[htb]
    \centering
	  \includegraphics[width=0.5\textwidth]{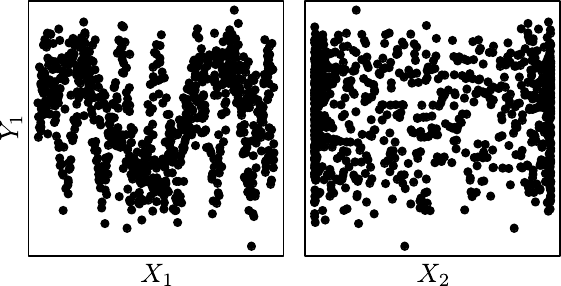}
	  \caption{The two pairs of margins of the sine mixture. In the left plot we see the superimposed sine signals, in the right plot the margins that determine the mixture.
  \label{fig:mix_sine_raw}}
\end{figure}

\begin{figure}[htb]
    \centering
	  \includegraphics[width=.7\textwidth]{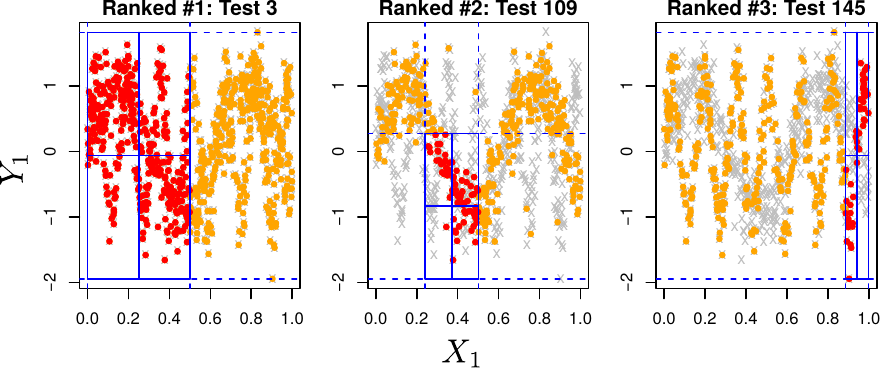}
	  \caption{Scatter plots for the observations in the three $2\times 2$ tables identified as most significant by {\tt MultiFIT} for the sine mixture scenario. The red points are observations that are within the cuboid that is tested, orange points are observations that are in a cuboid formed by expanding the tested cuboid so that the plotted margins are not subsetted. 
  \label{fig:mix_sine_top3}}
\end{figure}

{\bf Figure~\ref{fig:mix_sine_raw}} shows a simulated dataset of size 800. In the $(X_1,Y_1)$ margin (the left panel) we can see the superimposed sine waves. {\bf Figure~\ref{fig:mix_sine_top3}} shows three significant tables identified by \texttt{MultiFIT} using the same color coding technique from the previous example in which we can clearly discern between the different frequency waves.

\section{Application to a flow cytometry data set}
\label{sec:apps}
Flow cytometry is the standard biological assay used to measure single cell features known as markers, and is commonly used to quantify the relative frequencies of cell subsets in blood or disaggregated tissue. These features may be general physical, chemical or biological properties of a cell. Such data involve complex distributional features and are of massive sizes with typical sample sizes in the range of hundreds of thousands, which presents computational challenges to nonparametric data analytical tools.

For the evaluation, we used flow cytometry samples generated by an antibody panel designed to identify activated T cell subsets. We show the results of the dependency analysis on a single illustrative sample with 353,586 cells. For the analysis, we separated the markers into a vector of four `basic' markers (dump, CD3, CD4, CD8) and a vector of four `functional' markers (IFN, TNF, IL-2 and CD107). The basic markers are used in practice to first identify viable T cells by exclusion using the `dump' and CD3 markers, and then to further partition T cells into CD4-positive (`helper') and CD8-positive (`cytotoxic') subsets. The functional markers are used to identify the activation status of these T cell subsets and their functional effector capabilities (IL-2 is a T cell growth factor, IFN and TNF are inflammatory cytokines, and CD107 is a component of the mechanism used by T cells to directly kill infected and cancer cells).

We applied \texttt{MultiFIT} with Holm's multiple testing adjustment to the data to identifying dependency between the basic and functional markers. Our aim here is to demonstrate \texttt{MultiFIT}'s ability to handle such large data and to shed light on the underlying dependency, and so we ran the test exhaustively up to the maximal resolution of 4 - testing 102,416 $2\times2$ tables. The execution time of the algorithm in this setting is approximately 5 minutes on a laptop computer utilizing four 3.00GHz Intel\textsuperscript{\textregistered} Xeon(R) E3-1505M v6 CPU cores. 

\begin{figure}[p]
\centering
	\includegraphics[width=1\textwidth]{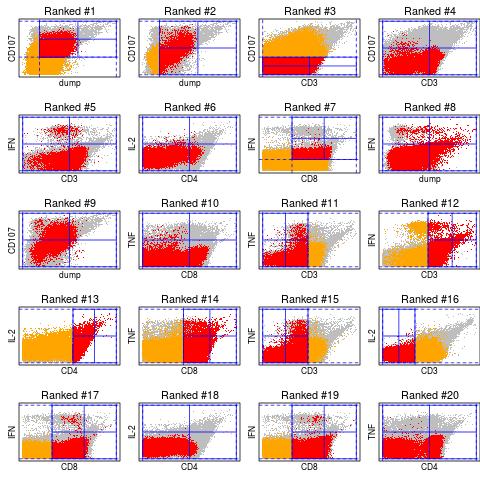}
\caption{Scatter plots of the observations identified by the 20 $2\times 2$ tables with the most significant $p$-values for the flow cytometry data set. Red indicates observations in the tested cuboid. Orange indicates observations in the same slice of the sample space, determined by the four markers other than the two margins plotted. Gray indicates the rest of the observations.\label{fig:flow_plots}}
\end{figure}

As the sample size is very large and the data clearly have strong marginal dependencies, \texttt{MultiFIT} identified hundreds of significant tests after multiple testing adjustment. Interested readers can run our code for this example in the Supplementary Materials to visualize the identified dependence structures. None of \texttt{HHG}, \texttt{DCov} and \texttt{dHSIC} was able to handle this amount of data and all ended in overflow errors. {\bf Figure~\ref{fig:flow_plots}} presents the visualization of the observations in the 20 $2\times 2$ table with the most significant $p$-values using the strategy described in Section~\ref{sec:rotated_circle}.

\section{Conclusion}
\label{sec:conclusion}
We have presented a scalable framework called \texttt{MultiFIT} for nonparametrically testing the independence of random vectors that achieves high computational scalability, decent statistical power, and the ability to shed light on the underlying dependency. We provide a finite-sample theoretical guarantee that \texttt{MultiFIT} controls the level exactly at any finite sample size without resorting to resampling or asymptotic approximation. The proposed approach is most suitable for multivariate problems up to tens of dimensions, can scale up to massive sample sizes, and thus can be useful for many modern data analyses. We have published an R package called {\tt MultiFit} on CRAN that implements the proposed method.

\section{Software}
    For the \texttt{MultiFIT} procedure we used our R package \texttt{MultiFit} on CRAN. For the Heller-Heller-Gorfine (\texttt{HHG}) test \citep{heller2013} we used the \texttt{HHG} package on CRAN. For Distance Covariance (\texttt{DCov}) \citep{szekely2009} we used the \texttt{energy} package on CRAN. For \texttt{dHSIC} \citep{pfister2018} we used the \texttt{dHSIC} package on CRAN.

\section{Acknowledgments}
The authors wish to thank the editor, an AE, and two referees for their valuable comments and suggestions. LM's research is partly supported by NSF grants DMS-1749789 and DMS-2013930. The flow cytometry data is collected through an EQAPOL collaboration with federal funds from the National Institute of Allergy and Infectious Diseases, National Institutes of Health, Contract Number HHSN272201700061C.

\newpage

\bibliographystyle{apalike}
\bibliography{manuscript.bib}

\newpage
\beginsupplement
\section*{Supplementary Material}
\section{Technical proofs}
\label{sec:supplement-proofs}
In this section, we establish proofs for Theorem~2.1, Theorem~2.2, and Corollary~2.1 in the main paper. We will introduce a number of definitions and lemmas along the way that will be used to complete the proofs.
\spacingset{1.45}

\begin{definition}\textit{Level-$k$ Canonical Marginal Partition}:\\
\label{def:partition}
$\mP^k$ is a \textit{level-$k$ canonical marginal partition} if
$
\mP^k = \left\{\left[\frac{l-1}{2^k},\frac{l}{2^k}\right)\right\}_{l\in\{1,...,2^k\}}.
$
\end{definition}
To simplify the notations in the proofs, we let $\bZ=(\bX,\bY)$. More specifically, we set $Z_1:=X_1,...,Z_{D_x}:=X_{D_x}$ and $Z_{D_x+1}:=Y_1,...,Z_D:=Y_{D_y}$. Thus, $\bZ$ is a random vector that distributed according to the distribution $F$, the joint sampling distribution of $(\bX,\bY)$.

\begin{figure}[!h]
\centering
\includegraphics[width=0.42\textwidth]{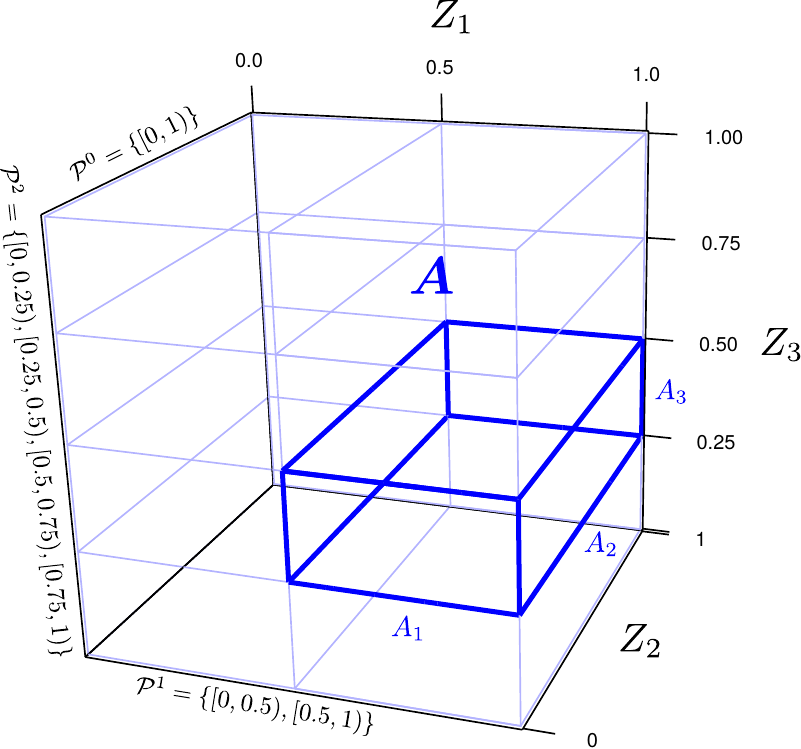}
\caption{Cuboid and stratum\label{fig:cuboid-strat-supplement}.\newline A 3D view of a $\bk$-stratum and the cuboid $A$ where $D=3$ and $\bk=(1,0,2)$. The thin blue lines delineate all $(1,0,2)$-cuboids, that is, the stratum  $\mA^{(1,0,2)}=\mP^{1}\times\mP^{0}\times\mP^{2}$. The thick blue lines delineate the cuboid $A$ for which $\bl=(2, 1, 2)$, i.e.,\ $A=A_1\times A_2\times A_3$ where $A_1=\left[\frac{2-1}{2^{1}},\frac{2}{2^{1}}\right)=[0.5,1)\in\mP^{1}$, $A_2=\left[\frac{1-1}{2^{0}},\frac{1}{2^{0}}\right)=[0,1)\in\mP^{0}$, and $A_3=\left[\frac{2-1}{2^{2}},\frac{2}{2^{2}}\right)=[0.25,0.5)\in\mP^{2}$. The resolution of this stratum is $3=1+0+2$.}
\end{figure}

For $\bk=(k_1,...,k_D)\in\N_0^D$, we refer to the partition $\P^{k_1}\times\cdots\times\P^{k_D}$ as the {\em $\bk$-stratum} of $\om$. Denote now any \textit{$\bk$-stratum} as $\mA^\bk$. As described in Section~\ref{sec:method}, we form $D$-dimensional cuboids by taking the Cartesian product of one interval from each of the $D$ canonical marginal partitions of $\om$. Given then $\bk=(k_1,...,k_D)\in\N_0^D$ a specific cuboid is determined by some $\bl=(l_1,...,l_D)$ with each $1\leq l_d \leq 2^{k_d}$ such that
$
A=\bigtimes_{d\in\D}\left[\frac{l_d-1}{2^{k_d}},\frac{l_d}{2^{k_d}}\right)
$. {\bf Figure~\ref{fig:cuboid-strat-supplement}} illustrates these definitions in a three dimensional space.

\begin{figure}[!h]
\centering
\includegraphics[width=0.42\textwidth]{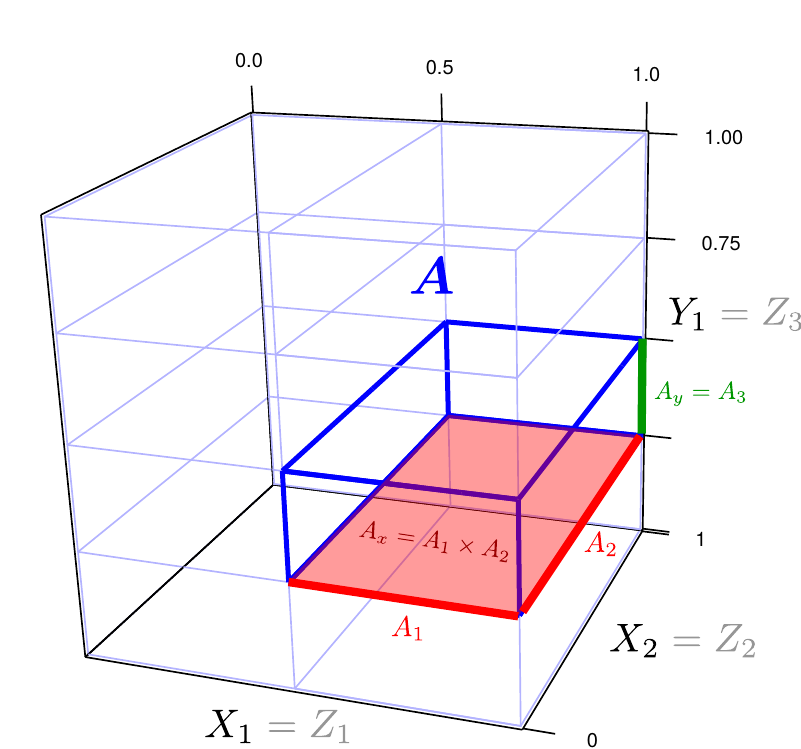}
\caption{$A_x$ and $A_y$\label{fig:cuboid-strat-Ax-Ay}.\newline A 3D view of a $\bk$-stratum and the cuboid $A$ where $D=3$, $\bk=(1,0,2)$ and $\bl=(2, 1, 2)$, the same as in {\bf Figure~\ref{fig:cuboid-strat-supplement}}. Here $D_x=2$ with $X_1=Z_1$ and $X_2=Z_2$; $D_y=1$ with $Y_1=Z_3$. The cuboid $A$ is represented now as $A=A_x\times A_y$ where $A_x=A_1\times A_2=\left[\frac{2-1}{2^{1}},\frac{2}{2^{1}}\right)\times\left[\frac{1-1}{2^{0}},\frac{1}{2^{0}}\right)=[0.5,1)\times [0,1)$, and $A_y=A_3=\left[\frac{2-1}{2^{2}},\frac{2}{2^{2}}\right)=[0.25,0.5)$.}
\end{figure}

We next define a discretized form of independence which we will show in Lemma~\ref{thm:thm1} fully characterizes the multivariate independence:
\begin{definition}\textit{$\bk$-independence}:\\
\label{def:mvkkindep}
For any $A\in\mA^\bk$, we can write $A=A_x \times A_y$ where
\begin{align*}
A_x =\bigtimes_{d=1}^{D_x}\left[\frac{l_d-1}{2^{k_d}},\frac{l_d}{2^{k_d}}\right)\text{ and }
A_y =\bigtimes_{d=D_x+1}^{D} \left[\frac{l_d-1}{2^{k_d}},\frac{l_d}{2^{k_d}}\right).
\end{align*}
({\bf Figure~\ref{fig:cuboid-strat-Ax-Ay}} illustrates the above notations in three dimensional space.)

We say that $\bX$ and $\bY$ are \textit{$\bk$-independent} and write it as $\bX \indepk \bY$ if for any $A\in \mA^{\bk}$, 
\begin{align*}
\Pb\left(\bX\in A_x, \bY\in A_y\right)=\Pb\left(\bX\in A_x\right)\cdot\Pb\left(\bY\in A_y\right).
\end{align*}
\end{definition}

\begin{figure}[!ht]
\centering
\includegraphics[width=0.42\textwidth]{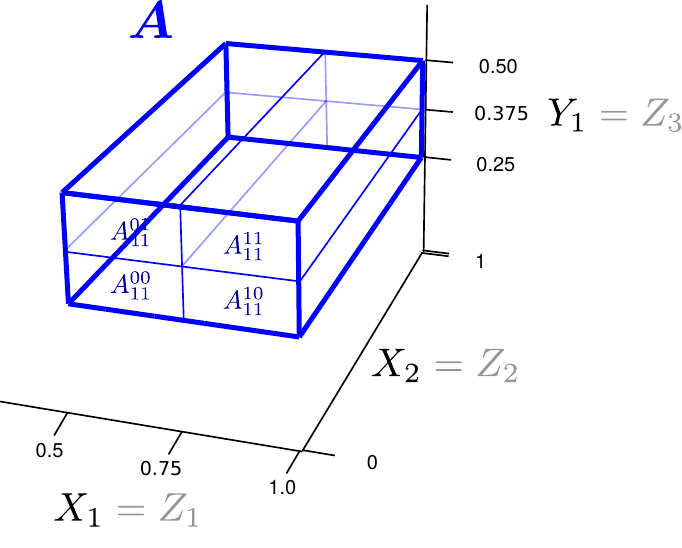}
\caption{($i=1$, $j=1$) blocks of $A$\label{fig:cuboid-strat-Aij-ab-supp}.\newline A 3D view of the $\bk$-cuboid $A$ where $D=3$, $\bk=(1,0,2)$ and $\bl=(2, 1, 2)$, the same as in {\bf Figures~\ref{fig:cuboid-strat-supplement}} and {\bf\ref{fig:cuboid-strat-Ax-Ay}}. Note that the $i=1$ dimension of $\bX$ corresponds to dimension 1 of $\bZ$ the $j=1$ dimension of $\bY$ corresponds to dimension 3 of $\bZ$. The blocks are:\newline
$A_{11}^{00}=\left[\frac{2\cdot 2-2+0}{2^{1+1}},\frac{2\cdot2-1+0}{2^{1+1}}\right)\times\left[\frac{1-1}{2^{0}},\frac{1}{2^{0}}\right)\times\left[\frac{2\cdot2-2+0}{2^{2+1}},\frac{2\cdot2-1+0}{2^{2+1}}\right)=[0.5,0.75)\times[0,1)\times[0.25,0.375)$,\newline
$A_{11}^{01}=\left[\frac{2\cdot 2-2+0}{2^{1+1}},\frac{2\cdot2-1+0}{2^{1+1}}\right)\times\left[\frac{1-1}{2^{0}},\frac{1}{2^{0}}\right)\times\left[\frac{2\cdot2-2+1}{2^{2+1}},\frac{2\cdot2-1+1}{2^{2+1}}\right)=[0.5,0.75)\times[0,1)\times[0.375,0.5)$,\newline
$A_{11}^{10}=\left[\frac{2\cdot 2-2+1}{2^{1+1}},\frac{2\cdot2-1+1}{2^{1+1}}\right)\times\left[\frac{1-1}{2^{0}},\frac{1}{2^{0}}\right)\times\left[\frac{2\cdot2-2+0}{2^{2+1}},\frac{2\cdot2-1+0}{2^{2+1}}\right)=[0.75,1)\times[0,1)\times[0.25,0.375)$,\newline
$A_{11}^{11}=\left[\frac{2\cdot 2-2+1}{2^{1+1}},\frac{2\cdot2-1+1}{2^{1+1}}\right)\times\left[\frac{1-1}{2^{0}},\frac{1}{2^{0}}\right)\times\left[\frac{2\cdot2-2+1}{2^{2+1}},\frac{2\cdot2-1+1}{2^{2+1}}\right)=[0.75,1)\times[0,1)\times[0.375,0.5)$.
}
\end{figure}

\begin{definition}\textit{$(i,j)$-blocks of a cuboid}:\\
\label{def:quadrants}
For every cuboid $A\in\mA^\bk$, $i\in\Dx$ and $j\in\Dy$, one can partition $A$ into four blocks by dividing $A$ in the $(i,j)$th face (that is, the side of $A$ spanned by the $i$th and $j$th dimensions) while keeping the other dimensions intact.
\[
A = A_{ij}^{00} \cup A_{ij}^{01} \cup A_{ij}^{10} \cup A_{ij}^{11},
\]
where for $a,b\in\{0,1\}$,
\begin{align*}
A_{ij}^{ab}=\bigtimes_{d=1}^D
\begin{cases}
\left[\frac{2l_d-2+a}{2^{k_{\!d}+1}},\frac{2l_d-1+a}{2^{k_{\!d}+1}}\right)&\text{if }d=i\\
\left[\frac{2l_d-2+b}{2^{k_{\!d}+1}},\frac{2l_d-1+b}{2^{k_{\!d}+1}}\right)&\text{if }d=D_x+j\\
\left[\frac{l_d-1}{2^{k_{\!d}}},\frac{l_d}{2^{k_{\!d}}}\right)&\text{if }d\in\D\setminus\{i,D_x+j\}\\
\end{cases}.
\end{align*}
\end{definition}
{\bf Figure~\ref{fig:cuboid-strat-Aij-ab-supp}} illustrates Definition~\ref{def:quadrants} in three dimensions.

Lemma~\ref{thm:thm1} establishes an equivalence between multivariate independence and a cascade of discretized multivariate independence relations:
\begin{lemma}
\label{thm:thm1}
\begin{align*}
\bX\indep\bY \slrs  \bX\indepk\bY \quad\text{for all $\bk\in\N_0^{D}$}.
\end{align*}
\end{lemma}

\begin{proof}
$\Rightarrow$: Immediate.\\
$\Leftarrow$:\\
Let $\mP_x=\bigcup_{\bk_{\{1,\ldots D_x\}}\in\N_0^{D_x}} \mP^{1}\times \ldots \mP^{D_x}$. Therefore $\sigma(\mP_x)=\mathcal{B}([0,1]^{D_x})$.\\ For $A_x\in\mathcal{B}([0,1]^{D_x})$ let 
$[\bX\in A_x]=\{\bm{\omega}: \bX(\bm\omega)\in A_x\}$, then $\sigma(\bX)=\{[\bX\in A_x], A_x\in \mathcal{B}([0,1]^{D_x})\}$.\\
Let $\mathcal{Q}_x=\bigcup_{\bk_{\{1,\ldots D_x\}}\in\{-1,0,1,2,...\}^{D_x}}\mathcal{Q}^{k_{1}}\times...\times\mathcal{Q}^{k_{D_x}}$ so that $\mathcal{Q}^{-1}:=\emptyset$ and $\forall k\geq0$, $\mathcal{Q}^k=\mP^k$. 

Let $\mathcal{C}_\bX=\{[\bX\in B_x], B_x\in \mathcal{Q}_x\}$.

Hence $\sigma(\mathcal{C}_\bX)=\sigma(\bX^{-1}(B_x), B_x\in\mathcal{Q}_x)=\sigma(\bX^{-1}(\mathcal{Q}_x))
=\bX^{-1}(\sigma(\mathcal{Q}_x))=\bX^{-1}(\sigma(\mP_x))=\sigma(\bX)$.\\
Note that $\mathcal{C}_\bX$ is a $\pi$-system:\\
$E,E'\in\mathcal{C}_\bX \srs E=\left[\bX\in \left(\emptyset\text{ or }\left[\frac{l_1-1}{2^{k_1}},\frac{l_1}{2^{k_1}}\right)\right)\times\ldots\times\left(\emptyset\text{ or }\left[\frac{l_{D_x}-1}{2^{k_{D_x}}},\frac{l_{D_x}}{2^{k_{D_x}}}\right)\right)\right]$ and\newline $E'=\left[\bX\in \left(\emptyset\text{ or }\left[\frac{l'_1-1}{2^{k'_1}},\frac{l'_1}{2^{k'_1}}\right)\right)\times\ldots\times\left(\emptyset\text{ or }\left[\frac{l'_{D_x}-1}{2^{k'_{D_x}}},\frac{l'_{D_x}}{2^{k'_{D_x}}}\right)\right)\right]$ for some $\bl, \bl', \bk, \bk'$.
Therefore:
\begin{align*}
E\cap E'&=\left[\bX\in \left\{\left(\emptyset\text{ or }\left[\frac{l_1-1}{2^{k_1}},\frac{l_1}{2^{k_1}}\right)\right)\times\ldots\times\left(\emptyset\text{ or }\left[\frac{l_{D_x}-1}{2^{k_{D_x}}},\frac{l_{D_x}}{2^{k_{D_x}}}\right)\right)\right\}\right.\bigcap\\&\hspace{5pc}\left.\left\{\left(\emptyset\text{ or }\left[\frac{l'_1-1}{2^{k'_1}},\frac{l'_1}{2^{k'_1}}\right)\right)\times\ldots\times\left(\emptyset\text{ or }\left[\frac{l'_{D_x}-1}{2^{k'_{D_x}}},\frac{l'_{D_x}}{2^{k'_{D_x}}}\right)\right)\right\}\right]\\
&=\left[\bX\in \left(\emptyset\text{ or }\left[\frac{l_1-1}{2^{k_1}},\frac{l_1}{2^{k_1}}\right)\cap \left[\frac{l'_1-1}{2^{k'_1}},\frac{l'_1}{2^{k'_1}}\right)\right)\times
\ldots\right.\\&\hspace{5pc}\left.\times\left(\left[\frac{l_{D_x}-1}{2^{k_{D_x}}},\frac{l_{D_x}}{2^{k_{D_x}}}\right)\cap \left[\frac{l'_{D_x}-1}{2^{k'_{D_x}}},\frac{l'_{D_x}}{2^{k'_{D_x}}}\right)\right)\right]\in\mathcal{C}_\bX
\end{align*}
Similarly, define $\sigma(\bY)$ and $\mathcal{C}_\bY$, another $\pi$-system, independent of $\mathcal{C}_\bX$ and $\sigma(\bY)=\sigma(\mathcal{C}_\bY)$. By the basic criterion, then, $\sigma(\bX)\indep \sigma(\bY)$.
\end{proof}

The following lemma shows that if $\bX$ and $\bY$ are $\bk$-independent, they are also independent on all coarser strata.
\begin{lemma}
\label{lem:lem2}
If $\bX \indepk \bY$ then $\bX \indepkp \bY$ for all $\bk'\leq\bk$.
\end{lemma}

\begin{proof}
\label{pf:thm2.2}
Let $F_\bX$ be the probability distribution of $\bX$ and $F_\bY$ be the probability distribution of $\bY$. Then $\bX\indepk\bY \slrs F(A)=F_\bX(A_x)F_\bY(A_y)$ for all $A\in\mA^\bk$.

It is enough to show that $\bX \indepk \bY \srs \bX \indepkp \bY$ for any $\bk'$ such that (i) $k'_i=k_i-1$ for some $i\in\{1,...,D_x\}$ and (ii) $k'_d=k_d$ for all $d\in\D\setminus\{i\}$.

Hence, assuming $\bX \indepk \bY$ we get for $A\in\mA^{\bk'}$ that:
\begin{align*}
F\left(A\right)
	&=
	F\left(A_1\times ...\times A_{i-1}\times A_i^0\times 
	A_{i+1}\times ...\times A_{D_x}\times A_y\right)\\
	&\hspace{1pc}
	+F\left(A_1\times ...\times A_{i-1}\times A_i^1\times 
	A_{i+1}\times ...\times A_{D_x}\times A_y\right)\\
	&=\FX\left(A_1\times ...\times A_{i-1}\times A_i^0\times 
	A_{i+1}\times ...\times A_{D_x}\right)\FY(A_y)\\
	&\hspace{1pc}
	+\FX\left(A_1\times ...\times A_{i-1}\times A_i^1\times 
	A_{i+1}\times ...\times A_{D_x}\right)\FY(A_y)\\
	&=\FX\left(A_x\right)
	\FY(A_y)
\end{align*}
As required.
\end{proof}

In Lemma~\ref{thm:thm2} we will show how to characterize the multivariate $\bk$-independence with a collection of univariate $(i,j)$ odds-ratios. However, before we state and prove Lemma~\ref{thm:thm2} we develop additional notations and provide discretized versions of some basic results in probability. Denote:

$\bd\subset\{1,..,D\}$, $\bi=\bd\cap\Dx$ and 
$\bj=\left\{j: j+D_x \in \bd\cap\{D_x+1,...,D\}\right\}$\\
And
\begin{align*}
\bX_\bi &= \{X_i: i\in\bi\}, &\bX_{(\bi)} &= \{X_{i'}: i'\in\Dx\setminus\bi\}\\
\bY_\bj &= \{Y_j: j\in\bj\},
& \bY_{(\bj)} &= \{Y_{j'}: j'\in\Dy\setminus\bj\}\\
\bZ_{\bd}&=\bX_{\bi}\times\bY_{\bj}=\{Z_{d}:d\in\bd\},&\bZ_{(\bd)}&=\bX_{(\bi)}\times\bY_{(\bj)}=\{Z_{d'}:d'\in\D\setminus\bd\}
\end{align*}
\begin{align*}
A_{x,\bd}&=\bigtimes_{d\in\bd\cap\Dx}A_d,& A_{x,(\bd)}&=\bigtimes_{d'\in\Dx\setminus\bd}A_{d'}\\A_{y,\bd}&=\bigtimes_{d\in\bd\cap\Dyd}A_d,
& A_{y,(\bd)}&=\bigtimes_{d'\in\Dyd\setminus\bd}A_{d'}\\
A_{\bd}&=A_{x,\bd}\times A_{y,\bd}=\bigtimes_{d\in\bd}A_d,&
A_{(\bd)}&=A_{x,(\bd)}\times A_{y,(\bd)}=\bigtimes_{d'\in\D\setminus\bd} A_{d'}
\end{align*}
\begin{align*}
\bk_{\bd} = \{k_{d}: d\in\bd\},\quad\quad 
\bk_{(\bd)}= \{k_{d'}: d'\in\D\setminus\bd\}
\end{align*}

\begin{definition}\textit{Conditional $\bk$-independence}:\\
\label{def:cdtl-kkindep}
We say that $\bX_\bi$ and $\bY_\bj$ are $\bk$-independent conditional on 
$\bZ_{(\bd)}$ and write it 
as $\bX_\bi \indepk \bY_\bj \mid \bZ_{(\bd)}$ if for any $A\in \mA^{\bk}$:
\begin{align*}
\Pb\huge(\bX_\bi\in A_{x,\bd},&\bY_\bj\in A_{y,\bd} \mid \bZ_{(\bd)}\in A_{(\bd)}\huge)=\Pb\huge(\bX_\bi\in A_{x,\bd}\mid \bZ_{(\bd)}\in A_{(\bd)}\huge)
	\cdot\Pb\huge(\bY_\bj\in A_{y,\bd}\mid \bZ_{(\bd)}\in A_{(\bd)}\huge)
\end{align*}
Or equivalently:
\begin{align*}
\Pb\huge(\bX_\bi\in A_{x,\bd} &\mid \bY_\bj\in A_{y,\bd}, \bZ_{(\bd)}\in A_{(\bd)}\huge)=\Pb\huge(\bX_\bi\in A_{x,\bd} \mid \bZ_{(\bd)}\in A_{(\bd)}\huge)
\end{align*}
\end{definition}

When $k_{d'}=0$ for some $d'\in\D\setminus\bd$, $\Omega_{Z_{d'}}=[0,1]$ for those indices and hence our notation may be compacted. For example, if $k_{d'}=0$ for all $d'\in\D\setminus\bd$:
\begin{align*}
\bX_\bi \indepk \bY_\bj | \bZ_{(\bd)} \slrs \bX_\bi \indepkdd \bY_\bj
\end{align*}

We next provide a discretized version of some basic results in probability:

\begin{lemma}Contraction:\\
For $\bd\subset\D$ such that $\Dx\subset \bd$ (i.e. $\bi=\Dx$, $\bX_\bi=\bX$ and $\bY_{(\bj)}=\bZ_{(\bd)}$):
\begin{align*}
\begin{cases}
\bX \indepk \bY_\bj \mid \bZ_{(\bd)}\\
\bX \indepkd \bY_{(\bj)}
\end{cases}
\srs 
\bX \indepk \bY
\end{align*}
\end{lemma}
\begin{proof} Immediate from definition.\end{proof}

\begin{lemma}Decomposition:\\
For $\bd\subset\D$ such that $\Dx\subset \bd$ (i.e. $\bi=\Dx$ and $\bX_\bi=\bX$):
\begin{align*}
\bX \indepk \bY
\srs 
\begin{cases}
\bX \indepkdd\bY_{\bj}\\
\bX \indepkd \bY_{(\bj)}
\end{cases}
\end{align*}
\end{lemma}
\begin{proof} Immediate from definition.\end{proof}

\begin{lemma}Weak Union:\\
For $\bd\subset\D$ such that $\Dx\subset \bd$ (i.e. $\bi=\Dx$, $\bX_\bi=\bX$, $\bY_{\bj}=\bZ_{\bd}$ and $\bY_{(\bj)}=\bZ_{(\bd)}$):
\begin{align*}
\bX \indepk \bY
\srs 
\begin{cases}
\bX \indepk \bY_\bj \mid \bZ_{(\bd)}\\
\bX \indepk \bY_{(\bj)}\mid \bZ_\bd
\end{cases}
\end{align*}
\end{lemma}
\begin{proof} Immediate from definition.\end{proof}

\begin{definition}
For $A_d\in\mP^{k_d-1}$ such that $A_d=\left[\frac{l_d-1}{2^{k_d-1}},\frac{l_d}{2^{k_d-1}}\right)$, define $A_d^0=\left[\frac{2l_d-2}{2^{k_d}},\frac{2l_d-1}{2^{k_d+1}}\right)\in\mP^{k_d}$ and $A_d^1=\left[\frac{2l_d-1}{2^{k_d+1}},\frac{2l_d}{2^{k_d+1}}\right)\in\mP^{k_d}$.
\end{definition}

\begin{definition}
\label{def:ij_coarser_strata}
Given $\bk\in\N_0^D$ and $i\in\Dx$ $j\in\Dy$, let
\[
\bk[i, j] = \left\{\bk'\in\N_0^D: k'_i<k_i, k'_{D_x+j}<k_{D_x+j} \text{ and } k'_d\leq k_d \text{ for all } d\in\D\setminus\{i,D_x+j\}\right\}
\]
and 
\[
\mA^{\bk[i, j]} = \bigcup_{\bk'\in\bk[i, j]}\mA^{\bk'}.
\]
\end{definition}

\begin{definition}
Given $\bk\in\N_0^D$ and $i\in\Dx$, $j\in\Dy$ let 
\begin{align*}
[k_i, k_j](\bk_{<i,<j}) = \huge\{\bk'\in\N_0^D: \quad&k'_i<k_i, k'_{D_x+j}<k_{D_x+j} \text{ and }\\&k'_d=k_d  \text{ for all } d\in\{1,...,i-1\}\cup\{D_x+1,...,D_x+j-1\}\\&\text{ and }
\\&k'_d=0  \text{ for all } d\in\{i+1,...,D_x\}\cup\{D_x+j+1,...,D\}\huge\}
\end{align*}
Accordingly, let
\[
\mA^{[k_i, k_j](\bk_{<i,<j})} = \bigcup_{\bk'\in[k_i, k_j](\bk_{<i,<j})}\mA^{\bk'},
\]
which denotes the totality of all cuboids in strata that are coarser than the stratum $\mA^{\bk}$ along the margins $i$ and $j$ such that margins $\{i+1,...,D_x\}$ and $\{D_x+j+1,...,D\}$ are allowed any value in $[0,1]$.
\end{definition}

Lemma~\ref{thm:thm2} ties the $\bk$-independence of the random vectors $\bX$ and $\bY$ with the $(i,j)$ odds-ratios:
\begin{lemma}
\label{thm:thm2} For any $\bk>\mathbf{0}_{D}$
\begin{align*}
\bX\indepk\bY\slrs \theta_{ij}(A)=1 \quad &\forall i \in \{1,\ldots,D_x\}, \hspace{.5pc} j \in \{1,\ldots,D_y\}, \hspace{.5pc} A\in \mA^{\bk'}\\ &\forall \text{$\bk'\in \N_0^D$ such that $\bk'\leq \bk$ with $k'_i<k_i$ and $k'_{D_x+j}<k_{D_x+j}$.} 
\end{align*}
\end{lemma}

\begin{proof}
$\Rightarrow$:\\
Assume $\bX\indepk\bY$ for some $\bk\in\N_0^D$.\\
Let $i\in\Dx$ and $j\in \Dy$.\\
By applying the weak union lemma twice we get that $X_i \indepkij Y_j \mid \bZ_{(\{i, D_x+j\})}$.\\
Notice that for $\bk'$ such that $k'_i=k_i-1$, $k'_{D_x+j}=k_{D_x+j}-1$, $k'_d=k_d$ for all $d\in\D\setminus\{i,D_x+j\}$ and $A\in\mA^{\bk'}$ we get that $A_{ij}^{00}, A_{ij}^{11}, A_{ij}^{01}, A_{ij}^{10} \in \mathcal{A}^{\bk}$. So $\theta_{ij}(A)=1$ by the definitions of conditional independence and the $(i,j)$ odds-ratios.\\
By Lemma~\ref{lem:lem2} we get that indeed $\theta_{ij}(A)=1$ for $A \in \mathcal{A}^{[k_i,k_j]}$.\\

$\Leftarrow$:\\
First, notice that it suffices to show that:
\begin{align*}
\bX\indepk\bY\sls &\theta_{ij}(A)=1
\\ &\forall i \in \{1,2,\ldots,D_x\}, \hspace{.5pc} j \in \{D_x+1,\ldots,D\}, \hspace{.5pc} A\in \mA^{[k_i, k_j](\bk_{<i,<j})}
\end{align*}
Since then we may rely on the opposite direction to get that $\theta_{ij}(A)=0$ for all $A\in \mathcal{A}^{[k_i,k_j]}$.\\

Examine \[X_i\indep_{\bk_{\{1,\ldots,i,D_x+1,\ldots,D_x+j\}}} Y_j\mid \bX_{\{1,\ldots,i-1\}}, \bY_{\{1,\ldots,j-1\}}\]

To see that the above is true, let $\bk'\in\N_0^D$ such that 
$k'_d=k_d  \text{ for all } d\in\{1,...,i\}\cup\{D_x+1,...,D_x+j\}$ and\newline $k'_d=0  \text{ for all } d\in\{i+1,...,D_x\}\cup\{D_x+j+1,...,D\}$.

For a given $A\in\mA^{\bk'}$ let $x,y$ be a pair of univariate random variables whose joint distribution is given by $G:=F_{X_i,Y_j\mid \bX_{\{1,\ldots,i-1\}}\in A_{x,{\{1,\ldots,i-1\}}}, \bY_{\{1,\ldots,j-1\}}\in A_{y,{\{D_x+1,\ldots,D_x+j-1\}}}}$. By assumption: 
\begin{align*}
1=\theta_{ij}(A)&=\frac{F(A_{ij}^{00})\cdot F(A_{ij}^{11})}{F(A_{ij}^{01})\cdot F(A_{ij}^{10})}\\&
=\frac{G(A_i^0\times A_{D_x+j}^0)\cdot G(A_i^1\times A_{D_x+j}^1)}{G(A_i^0\times A_{D_x+j}^1)\cdot G(A_i^1\times A_{D_x+j}^0)}
\end{align*}

Since the above holds for every $A\in\mA^{[k_i, k_j](\bk_{<i,<j})}$ we may apply Theorem 2 from \cite{mamao2017} and conclude that $x\indepkij y$ which is equivalent to $X_i\indep_{\bk_{\{1,\ldots,i,D_x+1,\ldots,D_x+j\}}} Y_j\mid \bX_{\{1,\ldots,i-1\}}, \bY_{\{1,\ldots,j-1\}}$.
\newpage
Next, examine:\\
\begin{align*}
(\star)&
\begin{cases}
X_1\indep_{\bk_{\{1,D_x+1\}}} Y_1\\
X_1\indep_{\bk_{\{1,D_x+1,D_x+2\}}} Y_2\mid Y_1\\
X_1\indep_{\bk_{\{1,D_x+1,D_x+2,D_x+3\}}} Y_3\mid \bY_{\{1, 2\}}\\
\hspace{4pc}\vdots\\
X_1\indep_{\bk_{\{1,D_x+1\ldots,D-1\}}} Y_{D_y-1}\mid \bY_{\{1,...,D_y-2\}}\\
X_1\indep_{\bk_{\{1,D_x+1\ldots,D\}}} Y_{D_y}\mid \bY_{\{1,...,D_y-1\}}
\end{cases}\\
(\star\star)&
\begin{cases}
X_2\indep_{\bk_{\{1,2,D_x+1\}}} Y_1\mid X_1\\
X_2\indep_{\bk_{\{1,2,D_x+1,D_x+2\}}} Y_2\mid X_1, Y_1\\
\hspace{4pc}\vdots\\
X_2\indep_{\bk_{\{1,2,D_x+1,\ldots D\}}} Y_{d_y}\mid X_1, \bY_{\{1,...,D_y-1\}}
\end{cases}\\
&\hspace{5pc}\vdots\\
(\star\star\star)&
\begin{cases}
X_{D_x}\indep_{\bk_{\{1,\ldots,D_x,D_x+1\}}} Y_1\mid \bX_{\{1,...,D_x-1\}}\\
X_{D_x}\indep_{\bk_{\{1,\ldots,D_x,D_x+1,D_x+2\}}} Y_2\mid \bX_{\{1,...,D_x-1\}}, Y_1\\
\hspace{4pc}\vdots\\
X_{D_x}\indepk Y_{D_y}\mid \bX_{\{1,...,D_x-1\}}, \bY_{\{1,...,D_y-1\}}
\end{cases}
\end{align*}
Each of the above rows is obtained by the previous argument. Applying the contraction lemma recursively from top to bottom to each of the rows in $(\star)$ shows that $X_1\indep_{\{1,D_x,\ldots,D\}}\bY$. Further applying the contraction lemma to the latter result and the rows of $(\star\star)$ shows that $\bX_{\{1,2\}}\indep_{\{1,2,D_x,\ldots,D\}}\bY$. And a similar application of the contraction lemma to the previous results and all the rows up to $(\star\star\star)$ shows that $\bX\indepk\bY$.
\end{proof}

Note, for every $i\in\Dx$ and $j\in\Dy$, and for each of the above conditions\\ 
$X_i\indep_{\bk_{\{1,\ldots,i-1,D_x+1,\ldots,D_x+j\}}} Y_j\mid \bX_{\{1,...,i-1\}}, \bY_{\{1,...,j-1\}}$
there are $(2^{k_{i}}-1)(2^{k_{j}}-1)$ one degrees of freedom tests required, each repeated due to the conditioning $\prod_{s=1}^{i-1} 2^{k_{s}}\cdot \prod_{t=D_x+1}^{D-1} 2^{k_{t}}$ times. Summing over $j$, we get that for each $i$, there are $(2^{k_{i}}-1)\cdot \prod_{s=1}^{i-1} 2^{k_{s}}\cdot (2^{\sum_{t=D_x+1}^{D} k_{t}}-1)$ one degrees of freedom tests. Summing those over $i$ we get that overall we need to perform $(2^{\sum_{s=1}^{D_x} k_{s}}-1)(2^{\sum_{t=D_x+1}^{D} k_{t}}-1)$ one degrees of freedom tests.\par
Under $H_1$ there are $2^{\sum_{s=1}^{D_x} k_{s}}\cdot2^{\sum_{t=D_x+1}^{D} k_{t}}-1$ degrees of freedom, under $H_0$ there are $(2^{\sum_{s=1}^{D_x} k_{s}}-1)+(2^{\sum_{t=D_x+1}^{D} k_{t}}-1)$ degrees of freedom, and thus we need $(2^{\sum_{s=1}^{D_x} k_{s}}-1)(2^{\sum_{t=D_x+1}^{D} k_{t}}-1)$ degrees of freedom to identify a difference between the null and the alternative. It follows from the proof that we may indeed use $(2^{\sum_{s=1}^{D_x} k_{s}}-1)(2^{\sum_{t=D_x+1}^{D} k_{t}}-1)$ 1-degree of freedom independence tests to do so.\\

\noindent\textbf{Proof of Theorem \ref{thm:thm0}:}\\Immediate from Lemmas~\ref{thm:thm1} and~\ref{thm:thm2}.\\

\noindent\textbf{Proof of Theorem \ref{lem:lem3}:} 
Let $A$ be a cuboid in resolution $r$, $i\in\{1,\hdots,D_x\}$, $j\in\{1,\hdots,D_y\}$ and $p_{ij}(A)$ the $p$-value that is determined by the table $\{n(A_{ij}^{00}), n(A_{ij}^{01}), n(A_{ij}^{10}), n(A_{ij}^{11})\}$. For any $r>R^*$, whether or not $A$ is selected in the \texttt{MultiFIT} procedure for testing is determined by the $p$-values observed on the collection of all potential ancestral cuboids of $A$. Without loss of generality we let $R^*=0$ to simplify notation. The general case requires trivial changes to the proof. A cuboid $A$ will be selected in \texttt{MultiFIT} if there exists a sequence of nested cuboids, or a lineage, $A_0\subset A_1\subset \cdots \subset A_r$ of resolution $0,1,\ldots,r$ respectively such that each $A_{k+1}$ is a child cuboid of $A_{k}$ in the $(i_k,j_k)$-face, and moreover, the $p$-value of the $(i_k,j_k)$-table of $A_{k}$ is less than the threshold $p^*$. As such, the event that a cuboid $A$ is in $\C^{(r)}$ is in the $\sigma$-algebra generated by the $2\times 2$-table counts $\bm{n}(\bar{A}_{ij})$ for all $(i,j)$ pairs and all sets $\bar{A}$ that can be an ancestor cuboid of $A$ along some lineage. 

Suppose the resolution-$r$ cuboid $A$ is in the $\mA^{\bk}$ stratum with $|\bk|=\sum_{d=1}^{D} k_d=r$. Also, let $r_x=\sum_{d=1}^{D_x}k_d$ and $r_y=\sum_{d=D_x+1}^{D} k_d$. 
Any potential ancestor cuboid of $A$, denoted by $\bar{A}$, is 
the union of several sets in $\mA^{\bk}$, and thus the $(i,j)$-table of $\bar{A}$, $\bm{n}(\bar{A}_{ij})$, for every $(i,j)$ pair, is determined exactly if we know the counts in all sets in $\mA^{\bk}$. 

For any positive integer $\rho$, we denote the collection of all level-$\rho$ marginal partitions of $\Omega_{\bX}$ as 
$$\widetilde{\bm\mP}_x^{\rho}=\left\{\mP^{k_1}\times \hdots\times\mP^{k_{D_x}}: \sum_{d=1}^{D_x}k_d=\rho \right\}$$
and the collection of all level-$\rho$ marginal partitions of $\Omega_{\bY}$ as $$\widetilde{\bm\mP}_y^{\rho}=\left\{\mP^{k_{D_x+1}}\times \hdots\times\mP^{k_{D}}: \sum_{d=D_x+1}^{D}k_d=\rho \right\}$$

Consider the following sequence of nested marginal partitions on $\Omega_{\bX}$, $\widetilde{\P}_{x}^1\subset \widetilde{\P}_{x}^2\subset \cdots \subset \widetilde{\P}_{x}^{r_x} \subset \widetilde{\P}_{x}^{r_x+1}\subset \widetilde{\P}_{x}^{r_x+2}\subset \cdots \widetilde{\P}_{x}^{r_x+D_x}$ (where $\widetilde{\P}_{x}^1\in\widetilde{\bm\P}_{x}^1, \cdots ,\widetilde{\P}_{x}^{r_x+D_x}\in\widetilde{\bm\P}_{x}^{r_x+D_x}$), such that
we first divide $\Omega_{X_1}$ $k_1$ times to get $\widetilde{\P}_{x}^{1},\hdots,\widetilde{\P}_{x}^{k_1}$, followed by dividing $\Omega_{X_2}$ $k_2$ times to get $\widetilde{\P}_{x}^{k_1+1},\hdots,\widetilde{\P}_{x}^{k_1+k_2}$, and so on an so forth until dividing $\Omega_{X_{D_x}}$ $k_{D_x}$ times to get $\widetilde{\P}_{x}^{r_x-k_{D_x}+1},\hdots,\widetilde{\P}_{x}^{r_x}$. Then divide $\Omega_{X_i}$ once to get $\widetilde{\P}_{x}^{r_x+1}$, and finally divide each of the other $D_x-1$ dimensions once in any order to get $\widetilde{\P}_{x}^{r_x+2},\ldots,\widetilde{\P}_{x}^{r_x+D_x}$.

In exactly the same manner, we can construct a sequence of nested marginal partitions of $\Omega_{\bY}$, $\widetilde{\P}_{y}^1\subset \widetilde{\P}_{y}^2\subset \cdots \subset \widetilde{\P}_{y}^{r_y} \subset \widetilde{\P}_{y}^{r_y+1}\subset \widetilde{\P}_{y}^{r_y+2}\subset \cdots \widetilde{\P}_{x}^{r_y+D_y}$ (where    $\widetilde{\P}_{y}^1\in\widetilde{\bm\P}_{y}^1, \cdots ,\widetilde{\P}_{y}^{r_y+D_y}\in\widetilde{\bm\P}_{y}^{r_y+D_y}$) such that we first divide $\Omega_{Y_1}$ $k_{D_x+1}$ times to get $\widetilde{\P}_{y}^{1},\hdots,\widetilde{\P}_{y}^{k_{D_x+1}}$, followed by dividing $\Omega_{Y_2}$ $k_{D_x+2}$ times to get $\widetilde{\P}_{y}^{k_{D_x+1}+1},\hdots,\widetilde{\P}_{y}^{k_{D_x+1}+k_{D_x+2}}$, and so on an so forth until dividing $\Omega_{Y_{D_y}}$ $k_{D}$ times to get $\widetilde{\P}_{y}^{r_y-k_{D}+1},\hdots,\widetilde{\P}_{y}^{r_y}$. Then divide $\Omega_{Y_j}$ once to get $\widetilde{\P}_{y}^{r_y+1}$, and finally divide each of the other $D_y-1$ dimensions once in any order to get $\widetilde{\P}_{y}^{r_x+1},\ldots,\widetilde{\P}_{x}^{r_x+D_y}$.

Under these two marginal partition sequences, we have $A\in \widetilde{\P}_{x}^{r_x} \times \widetilde{\P}_{y}^{r_y}=\mA^{\bk}$, whereas the four child cuboids of $A$ with respect to the $(i,j)$-face are in the two strata $\widetilde{\P}_{x}^{r_x+1} \times \widetilde{\P}_{y}^{r_y}$ and $\widetilde{\P}_{x}^{r_x} \times \widetilde{\P}_{y}^{r_y+1}$. Moreover, any $(i,j)$-face of any ancestral cuboid of $A$ are formed by unions of sets that are not in the strata $\widetilde{\P}_{x}^{r_x+i'} \times \widetilde{\P}_{y}^{r_y+j'}$ for $i'=1,2,\ldots,D_x$ and $j'=1,2,\ldots,D_y$.

Now by Theorem~3 in \cite{mamao2017}, conditional on the $\bX$ and $\bY$ marginal values of the observations, the counts in any $(i,j)$-table of $A$ given the corresponding row and column totals are independent of the $\sigma$-algebra generated by all counts in the sets that are not of the form $\widetilde{\P}_{x}^{r_x+i'} \times \widetilde{\P}_{y}^{r_y+j'}$ for $i'=1,2,\ldots,D_x$ and $j'=1,2,\ldots,D_y$. Therefore, the counts in any $(i,j)$-table of $A$ are also independent of the $\sigma$-algebra generated by the $2\times 2$-table counts $\bm{n}(\bar{A}_{ij})$ for all sets $\bar{A}$ that can be ancestor cuboids of $A$ along some lineage and all $(i,j)$ pairs, and hence are also independent of the selection under the \texttt{MultiFIT} procedure. This completes the proof.  \hfill \qedsymbol

\vspace{1em}
\noindent\textbf{Proof of Corollary \ref{cor:cor1}:}
This corollary follows immediately since the $p$-value on the $(i,j)$-table of a cuboid $A$ is determined from the (central) hypergeometric distribution given the row and column totals of that table, which due to Theorem~\ref{lem:lem3}, is the actual sampling distribution of the table given the row and column totals under the null hypothesis of independence whether or not one conditions on the event that $A$ is selected for testing in the \texttt{MultiFIT} procedure.\hfill\qedsymbol\\

\noindent\textbf{Proof of Theorem \ref{thm:local_consistency}:
We adopt here a similar strategy to what was used in \cite[Theorem~5]{mamao2017}. We recall that the null distribution of the $2\times2$ contingency table on a cuboid $A$ arising from an i.i.d. sample conditional on its margins is the central hypergeometric distribution (as discussed in Section~\ref{sec:theory}) and the same table when the null does not hold is distributed according to the noncentral hypergeometric distribution. \cite{kou:1996} developed a normal approximation for the noncentral hypergeometric distribution (and accompanying results) on which we rely.}\\

Under the conditions in the theorem's statement, we can take $R_{max}=R^*$ -- that is, exhaustive testing. According then to our Strategy I for multiple testing adjustment (the holistic approach to multiple testing) with Bonferroni's correction 
\[\alpha_{ij}(A)=\alpha / \left(\sum_{\rho=0}^{R_{max}}  D_x\cdot D_y\cdot 2^\rho\cdot {{\rho+D-1}\choose{D-1}}\right)\]
is the table-specific threshold. According to our Strategy II for multiple testing adjustment (the resolution-specific approach to multiple testing) with exhaustive testing  up to resolution $R_{max}$ and utilizing Bonferroni's correction within resolution as well as between resolutions
\[\alpha_{ij}(A)=\alpha / \left({D_x\cdot D_y \cdot 2^r\cdot {{r+D-1}\choose{D-1}}\cdot R_{max}}\right)\] 
is the table-specific threshold. The selection of Bonferron's correction here ensures that the proof is valid for any correction that is less conservative -- e.g. Holm's correction that we use in our implementation.\\

We start by proving the theorem in a simplified case where we assume that $R^*=R_{max}$ is fixed but large enough such that $r\leq R^*$. It suffices to notice that under both instances $\alpha_{ij}(A)$ is constant when $R^*$ is constant. Let then $\theta_{ij}(A)=\theta$. Without loss of generality, let us assume that $\theta>1$.\\

By Theorem~2.2 in \cite{kou:1996}, we have that given $n(A^{0\cdot}_{ij}),n(A^{\cdot 0}_{ij}),n(A)$,
\[
Z_{n,\theta,i,j}(A) =\frac{n(A^{00}_{ij})-{\rm E}_{\theta} [n(A^{00}_{ij})\,|\,n(A^{0\cdot}_{ij}),n(A^{\cdot 0}_{ij}),n(A)]}{{\rm Var}^{1/2}_{\theta}[n(A^{00}_{ij})\,|\,n(A^{0\cdot}_{ij}),n(A^{\cdot 0}_{ij}),n(A)]} \rightarrow_{\mathcal{L}} {\rm N}(0,1).
\]
$p_{ij}(A)$, the $p$-value for the table determined by $n(A^{00}_{ij})$ when the margins $\{n(A^{0\cdot}_{ij}),n(A^{\cdot 0}_{ij}),n(A)\}$ are given, is computed for the two sided version of Fisher's exact test by summing the probabilities of all tables that are more extreme than the given one (that is, those with a smaller probability of occurrence compared to the table for which the $p$-value is computed according to the central hypergeometric distribution).\\

We can utilize the form $Z_{n,\theta,i,j}$ to write the probability of rejection for the test on the $(i,j)$-table of $A$. If $n(A^{00}_{ij})$ is less than or equal to the mode of the hypergeometric distribution with the parameters $\{n(A^{0\cdot}_{ij}),n(A^{\cdot 0}_{ij}),n(A)\}$ we have
\begin{align*}
{\rm P}&(p_{ij}(A)< \alpha_{ij}(A)\,|\,\theta_{ij}(A)=\theta,\bn_{r_x,0},\bn_{0,r_y})\\ 
>&{\rm P}\left(Z_{n,1,i,j}(A)>F_{A,n}^{-1}(\alpha_{ij}(A))\,|\,\theta_{ij}(A)=\theta,\bn_{r_x,0},\bn_{0,r_y}\right)
\end{align*}
where $F_{A,n}$ denotes the exact cdf of $Z_{n,1,i,j}$ given the marginal totals. Else, if $n(A^{00}_{ij})$ is greater than the mode of the hypergeometric distribution with the parameters $\{n(A^{0\cdot}_{ij}),n(A^{\cdot 0}_{ij}),n(A)\}$
\begin{align*}
{\rm P}&(p_{ij}(A)< \alpha_{ij}(A)\,|\,\theta_{ij}(A)=\theta,\bn_{r_x,0},\bn_{0,r_y})\\ 
>&{\rm P}\left(Z_{n,1,i,j}(A)>F_{A,n}^{-1}(1-\alpha_{ij}(A))\,|\,\theta_{ij}(A)=\theta,\bn_{r_x,0},\bn_{0,r_y}\right).
\end{align*}
Without loss of generality we continue to work with the second case.
\\


Assume now that indeed $\theta_{ij}(A)=\theta \ne1$. Then
\begin{align*}
\lim_{n\to\infty}&{\rm P}\left(Z_{n,1,i,j}(A) > F_{A,n}^{-1}(1-\alpha_{ij}(A))\,|\,\theta_{ij}(A)=\theta,\bn_{r_x,0},\bn_{0,r_y}\right)\\
&= \lim_{n\to\infty} {\rm P}\left(Z_{n,\theta,i,j}(A) > c_{n}F_{A,n}^{-1}(1-\alpha_{ij}(A)) - c_n d_n\,\Big|\,\theta_{ij}(A)=\theta,\bn_{r_x,0},\bn_{0,r_y}\right)
\end{align*}
where 
\[
c_{n}=\frac{{\rm Var}^{1/2}_{1}[n(A^{00}_{ij})\,|\,n(A^{0\cdot}_{ij}),n(A^{\cdot 0}_{ij}),n(A)]}{{\rm Var}^{1/2}_{\theta}[n(A^{00}_{ij})\,|\,n(A^{0\cdot}_{ij}),n(A^{\cdot 0}_{ij}),n(A)]}\]
and
\[
d_{n}=\frac{{\rm E}_{\theta} [n(A^{00}_{ij})\,|\,n(A^{0\cdot}_{ij}),n(A^{\cdot 0}_{ij}),n(A)] - {\rm E}_{1} [n(A^{00}_{ij})\,|\,n(A^{0\cdot}_{ij}),n(A^{\cdot 0}_{ij}),n(A)]}{{\rm Var}^{1/2}_{1}[n(A^{00}_{ij})\,|\,n(A^{0\cdot}_{ij}),n(A^{\cdot 0}_{ij}),n(A)]}. 
\]
By Corollary~2.1 in \cite{kou:1996} we have $1/\sqrt{\max(\theta, \theta^{-1})} \leq c_n \leq 1/\sqrt{\min(\theta, \theta^{-1})}$ for all $n$ and $d_n \asymp \sqrt{n}$ with $F^{\infty}$ probability 1.\\

By the above normal approximation $F_{A,n}^{-1}(1-\alpha_{ij}(A))\rightarrow \Phi^{-1}(1-\alpha_{ij}(A))$.\\

Therefore $c_{n}\Phi^{-1}(1-\alpha_{ij}(A)) - c_n d_n\rightarrow -\infty$ with $F^{\infty}$ probability 1, and so with $F^{\infty}$ probability 1,
\[
\lim_{n\to\infty} {\rm P}(p_{ij}(A) < \alpha_{ij}(A)\,|\,\bn_{r_x,0},\bn_{0,r_y}) = 1.
\]
This completes the proof for the case where $R^*$ is constant since the null of independence is rejected whenever $p_{ij}(A)<\alpha_{ij}(A)$ for some $A$, $i$ and $j$.\\

Assume now that $R_{max}=R^*=o(\log n)$. Under both the holistic and resolution specific strategies for multiple testing we have $\sqrt{n}\cdot \alpha_{ij,n}(A)\to\infty$ as $n\to\infty$.\\

Since $c_n d_n=O(\sqrt{n})$ with $F^{\infty}$ probability 1, we need to show that $c_n F_{A,n}^{-1}(1-\alpha_{ij}(A))=o(\sqrt{n})$ with $F^{\infty}$ probability 1 in order to establish consistency. Rewrite:
\begin{align*}
c_n F_{A,n}^{-1}&(1-\alpha_{ij,n}(A)) =\\& c_n\left(F_{A,n}^{-1}(1-\alpha_{ij,n}(A))-\Phi^{-1}(1-\alpha_{ij,n}(A))\right) + c_n\Phi^{-1}(1-\alpha_{ij,n}(A))
\end{align*}
Examine $c_n\left(F_{A,n}^{-1}(1-\alpha_{ij,n}(A))-\Phi^{-1}(1-\alpha_{ij,n}(A))\right)$: by Theorem~2.3 of \cite{kou:1996}, we have with $F^{\infty}$ probability 1,
\[
|\Phi(F_{A,n}^{-1}(1-\alpha_{ij,n}(A)) - (1-\alpha_{ij,n}(A))| < \gamma/\sqrt{n}
\]
for some positive constant $\gamma$ and large enough $n$. Therefore
\[
\Phi^{-1}(1-\alpha_{ij,n}(A)-\gamma/\sqrt{n}) < F_{A,n}^{-1}(1-\alpha_{ij,n}(A)) < \Phi^{-1}(1-\alpha_{ij,n}(A) + \gamma/\sqrt{n}).
\]
Since $1-\Phi(x) \asymp e^{-x^2/2}/x$ as $x\rightarrow \infty$ and  
$\sqrt{n}\cdot \alpha_{ij,n}(A)\to\infty$ as $n\to\infty$, we have
\[|\Phi^{-1}(1-\alpha_{ij,n}(A)-\gamma/\sqrt{n})-\Phi^{-1}(1-\alpha_{ij,n}(A))|\rightarrow 0\] and \[|\Phi^{-1}(1-\alpha_{ij,n}(A) + \gamma/\sqrt{n})-\Phi^{-1}(1-\alpha_{ij,n}(A))|\rightarrow 0\]
Hence,
\[
|F_{A,n}^{-1}(1-\alpha_{ij,n}(A))-\Phi^{-1}(1-\alpha_{ij,n}(A))| \rightarrow 0.\]

Examine $c_n\Phi^{-1}(1-\alpha_{ij,n}(A))$: since $1-\Phi(x) \asymp e^{-x^2/2}/x$ as $x\rightarrow \infty$ and $\sqrt{n}\cdot \alpha_{ij,n}(A)\to\infty$ as $n\to\infty$, we have $\Phi^{-1}(1-\alpha_{ij,n}(A))=o(\sqrt{n})$.

Therefore, we have with $F^{\infty}$ probability~1 that $c_n F_{A,n}^{-1}(1-\alpha_{ij}(A))=o(\sqrt{n})$. I.e., we have with probability~1 $p_{ij}(A)<\alpha_{ij}(A)$ for some $A$, $i$ and $j$ and therefore the null of independence is rejected with probability~1.
\hfill\qedsymbol\\

\newpage
\section{Pseudo-code for the {\tt MultiFIT} procedure}
\label{sec:algo_supp}
\begin{algorithm}[ht]
\caption{\texttt{MultiFIT} procedure for testing multivariate independence}
\label{alg:multifit}
\begin{algorithmic}
\\
\State{Let $\C^{(r)}$ be the collection of all cuboids of resolution $r$ for $r=0,1,2,\ldots,R^*$, and let $\C^{(r)}=\emptyset$ for $r=R^*+1,\ldots, R_{max}$.} \Comment{Step 0: Initialization}
\vspace{0.5em}

\For{$r$ in $0,1,2,\ldots,R_{max}$}  \Comment{For each resolution}
\For{each $A\in \C^{(r)}$} \Comment{For each cuboid selected for testing}
\For{$i$ in $1,2,\ldots,D_x$}
\For{$j$ in $1,2,\ldots,D_y$}
\State{Apply Fisher's exact test on the $(i,j)$-table of $A$ and record the $p$-value}\\
\Comment{Step 1a: Independence testing}

\If {$R^*\leq r < R_{max}$} 
    \If{the $(i,j)$-table of $A$ has a $p$-value smaller than a threshold $p^*$}
        \State{Add the four half cuboids of $A$ into $\C^{(r+1)}$}\\
            \Comment{Step 1b: Select cuboids for testing in the next resolution}
    \EndIf
\EndIf
\EndFor
\EndFor
\EndFor
\EndFor
\vspace{0.5em}
\State{Apply a multiple testing procedure that provides strong FWER control based on the recorded $p$-values.} \Comment{Step 2: Multiple testing control}
\end{algorithmic}
\end{algorithm}
\newpage 

\section{Scenarios for the power study}
\begin{table}[!ht]
\caption{Simulation Scenarios\label{tbl:scenarios}}
{\def\arraystretch{2}\tabcolsep=12pt
\begin{tabular}{c|c|c|c}
Scenario & \makecell{\# of Data\\Points} & \makecell{Max\\Res} & Simulation Setting \\\hline
Sine & 300 & 4 & \makecell{$X_1=Z$, $Y_1=Z'$, $X_2=U$,\\ $Y_2=\sin(5\pi\cdot X_2)+4\epsilon$}\\\hline
Circular & 300 & 4 & \makecell{$X_1=Z$, $Y_1=Z'$,$\theta\sim \mathrm{Uniform}(-\pi,\pi)$\\ $X_2=\cos(\theta)+\epsilon$, $Y_2=\sin(\theta)+\epsilon'$}\\\hline
Checkerboard & 1500 & 5 & \makecell{$W\sim \text{Multi-Bern}(\{1,2,3,4,5\},(1/5,1/5,1/5,1/5,1/5))$\\$V_1\sim \text{Multi-Bern}(\{1,3,5\},(1/3,1/3,1/3))$\\$V_2\sim \text{Multi-Bern}(\{2,4\},(1/2,1/2))$\\$X_1=Z$, $Y_1=Z'$, $X_2=W+\epsilon$,\\ $Y_2=\begin{cases}V_1+\epsilon', &\text{if $W$ is odd}\\V_2+\epsilon', &\text{if $W$ is even}\end{cases}$}\\\hline
Linear & 300 & 4&\makecell{$X_1=Z$, $Y_1=Z'$, $X_2=U$,\\ $Y_2=X_2+3\epsilon$}\\\hline
Parabolic & 300 &4& \makecell{$X_1=Z$, $Y_1=Z'$, $X_2=U$,\\ $Y_2=(X_2-0.5)^2+0.75\epsilon$}\\\hline
Local & 1000 & 6& \makecell{$X_1=Z$, $Y_1=Z'$, $X_2=Z''$,\\ $Y_2=\begin{cases}X_2+1/6\cdot\epsilon &\text{if }0<X_2, Z'''<0.7 \\ Z''', &\text{otherwise}\end{cases}$}\\\hline
\end{tabular}\par}
\begin{flushleft}
Six simulation scenarios. In all cases, $Z, Z', Z'', Z'''$ are i.i.d $\mathrm{N}(0,1)$. At each noise level $l=1,2,...,20$, $\epsilon, \epsilon'$ and $\epsilon''$ are i.i.d $\mathrm{N}(0,(l/20)^2)$, and $U\sim \mathrm{Uniform(0,1)}$. The maximal resolution is the algorithm's default: $\lfloor\log_2(n/10)\rfloor$ where $n$ is the number of data points. 
\end{flushleft}
\end{table}

\begin{table}[!ht]
\caption{``Spread'' Simulation Scenarios (part 1)\label{tbl:scenarios_spread}}
{\def\arraystretch{2}\tabcolsep=12pt
\begin{tabular}{c|c|c|c}
Scenario & \makecell{\# of Data\\Points} & \makecell{Max\\Res} & Simulation Setting \\\hline
Sine & 300 & 4 & \makecell{$X_1=U, X_2=U$\\
$Y_1 = \sin(5\pi X_1) + \cos(5\pi X_2) + 4\epsilon$\\
$Y_2 = - \sin(5\pi X_2) + \cos(5\pi X_1) + 4\epsilon'$
}\\\hline
Circular & 300 & 4 & \makecell{$Z_1, Z_2 \sim Beta(0.9, 0.27)$\\
$b_1, b_2, b_3, b_4 \sim \text{Multi-Bern}(\{-1,1\},(0.5, 0.5))$\\
$X_1 = b_1Z_1 + \epsilon$, $X_2 = b_2Z_2 + \epsilon'$\\
$Y_1 = b_3\sqrt{1-(0.2 b_1Z_1 + 0.8 b_2Z_2)^2} + \epsilon''$\\
$Y_2 = b_4\sqrt{1-(0.8 b_1Z_1 + 0.2 b_2Z_2)^2} + \epsilon'''$
}\\\hline
Checkerboard & 1500 & 5 & \makecell{$W, W' \sim \text{Multi-Bern}(\{1, 2, 3, 4, 5\}, (0.2,0.2,0.2,0.2,0.2))$\\
$X_1 = W + 0.75^2\epsilon$, $X_2 = W' + 0.75^2\epsilon'$\\
$m_1 \sim \text{Multi-Bern}(\{1,2\}, (0.2, 0.8))$\\
$m_2 \sim \text{Multi-Bern}(\{1,2\}, (0.8, 0.2))$\\
$V_1, V_1'\sim\text{Multi-Bern}(\{1,3,5\}, (1/3, 1/3, 1/3))$\\
$V_2, V_2' \sim \text{Multi-Bern}(\{2,4\}, (0.5, 0.5))$\\
$Y_1=\begin{cases}V_1+0.75^2\epsilon'', &\text{if $x_{m_1}$ is odd}\\V_2+0.75^2\epsilon'', &\text{if $x_{m_1}$ is even}\end{cases}$\\
$Y_2=\begin{cases}V_1'+0.75^2\epsilon''', &\text{if $x_{m_2}$ is odd}\\V_2'+0.75^2\epsilon''', &\text{if $x_{m_2}$ is even}\end{cases}$}\\\hline
\end{tabular}\par}
$\epsilon, \epsilon', \epsilon''$ and $\epsilon'''$ are i.i.d $\mathrm{N}(0,(l/20)^2)$, and $U\sim \mathrm{Uniform(0,1)}$. The maximal resolution is the algorithm's default: $\lfloor\log_2(n/10)\rfloor$ where $n$ is the number of data points.
\end{table}

\begin{table}[!ht]
\caption{``Spread'' Simulation Scenarios (part 2)\label{tbl:scenarios_spread2}}
{\def\arraystretch{2}\tabcolsep=12pt
\begin{tabular}{c|c|c|c}
Scenario & \makecell{\# of Data\\Points} & \makecell{Max\\Res} & Simulation Setting \\\hline
Linear & 300 & 4&\makecell{$X_1, X_2 \sim U(0,1)$\\
$Y_1 = X_1 - 2X_2 + 6\epsilon$,
$Y_2 = -X_1 + X_2 + 6\epsilon'$}\\\hline
Parabolic & 300 &4& \makecell{$X_1, X_2 \sim \mathrm{Uniform}(0,1)$\\
$Y_1 = (X_1 - 0.5)^2 - (X_2 - 0.5)^2 + 0.75\epsilon$\\
$Y_2 = (X_2 - 0.5)^2 - (X_1 - 0.5)^2 + 0.75\epsilon'$
}\\\hline
Local & 1000 & 6& \makecell{A spread linear signal scaled to (0,0.7) for all margins:\\
$x_1, x_2 \sim U(0,0.7)$\\
$y_1 = 0.7/2.1\cdot(x_1 - x_2 + 1.4)+1/6\cdot\epsilon$\\
$y_2 = 0.7/1.4\cdot(-x_1 + x_2 + 0.7)+1/6\cdot\epsilon'$\\
Embedded within a small portion of the space:\\
$X_1=Z$, $X_2=Z'$\\
$Y_1=\begin{cases}y_1 &\text{if }0<X_1, Z''<0.7 \\ Z'', &\text{otherwise}\end{cases}$\\
$Y_2=\begin{cases}y_3 &\text{if }0<X_2, Z'''<0.7 \\ Z'', &\text{otherwise}\end{cases}$
}\\\hline
\end{tabular}\par}
\begin{flushleft}
$Z, Z', Z'', Z'''$ are i.i.d $\mathrm{N}(0,1)$. At each noise level $l=1,2,...,20$, $\epsilon, \epsilon'$ and $\epsilon''$ are i.i.d $\mathrm{N}(0,(l/20)^2)$, and $U\sim \mathrm{Uniform(0,1)}$. The maximal resolution is the algorithm's default: $\lfloor\log_2(n/10)\rfloor$ where $n$ is the number of data points.
\end{flushleft}
\end{table}

\clearpage
\section{Numerical validation of level control through simulations\label{sec:fwer}}
\begin{figure}[ht]
    \centering
    \includegraphics[width=1\textwidth]{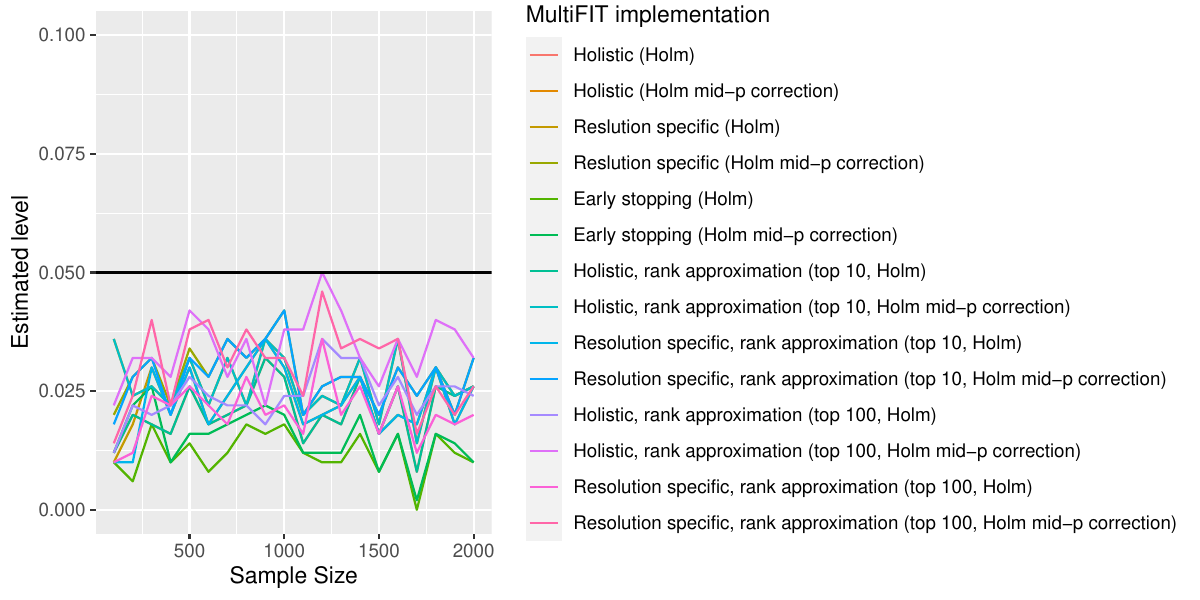}
    \caption{Estimated level versus sample size}
    \label{fig:fwer}
\end{figure}

To demonstrate that \texttt{MultiFIT} properly controls the level we executed 500 simulations with the default tuning parameters for various sample sizes. The underlying data $\{X_{1i}\}$, $\{X_{2i}\}$, $\{Y_{1i}\}$ and $\{Y_{2i}\}$ are drawn independently from a standard normal distribution for $i\in\{1,\hdots,n\}$ with $n \in \{100,200,\hdots,2000\}$. \textbf{Figure~\ref{fig:fwer}} shows the estimated level for \texttt{MultiFIT} with different variations for the independence tests on each table and multiple testing adjustment options on Fisher's exact test with mid-$p$ corrected $p$-values (see \cite{agresti2007} for a discussion on the mid-$p$ correction).

The results confirm the theoretical guarantees that the level can be controlled at any given level $\alpha$. In fact, the procedure appears to be a bit conservative in controlling the level. Note, although $\bX$ and $\bY$ are independent under the null hypothesis, the dependency structure between the different margins of $\bX$ is arbitrary, as well as the dependency structure between the different margins of $\bY$. Therefore, this simulation is not exhaustive. However, we repeated the estimation of the level under various dependency structures for the margins, and the results are consistent with these that are shown here.

\section{Scaling: comparison of scenarios for {\tt MultiFIT}}
\begin{figure}[ht]
    \centering
    \includegraphics[width=1\textwidth]{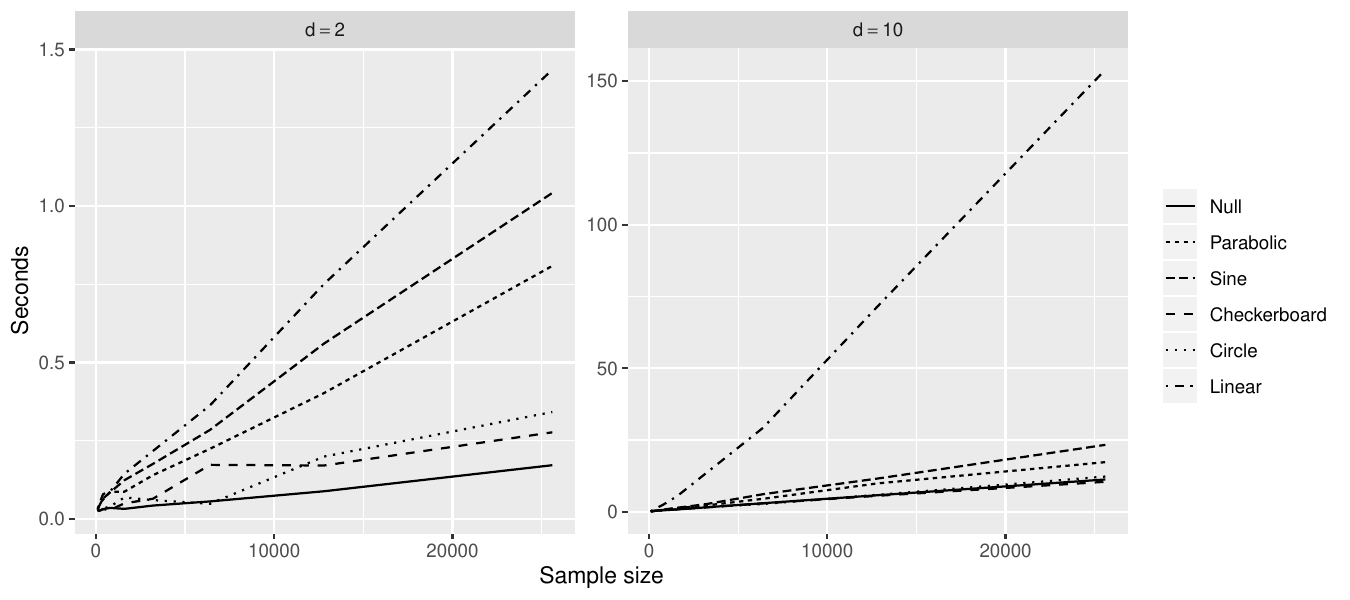}
\caption{Computational scalability: run-time vs. sample size for the six simulation scenarios from {\bf Table~\ref{tbl:scenarios}} when fitted with \texttt{MultiFIT} in different dimensionalities with $D_x=D_y=d$, $R^*=1$ and $l=3$. In all cases the ``linear'' scenario requires the most computations and the ``null'' scenario the least.\label{fig:scaling_supp}}
\end{figure}

\section{Power study: sensitivity analysis for the parameters $p^*$ and $R^*$ of {\tt MultiFIT}\label{sec:sensitivity}}
\begin{figure}[ht]
    \centering
    \includegraphics[width=1\textwidth]{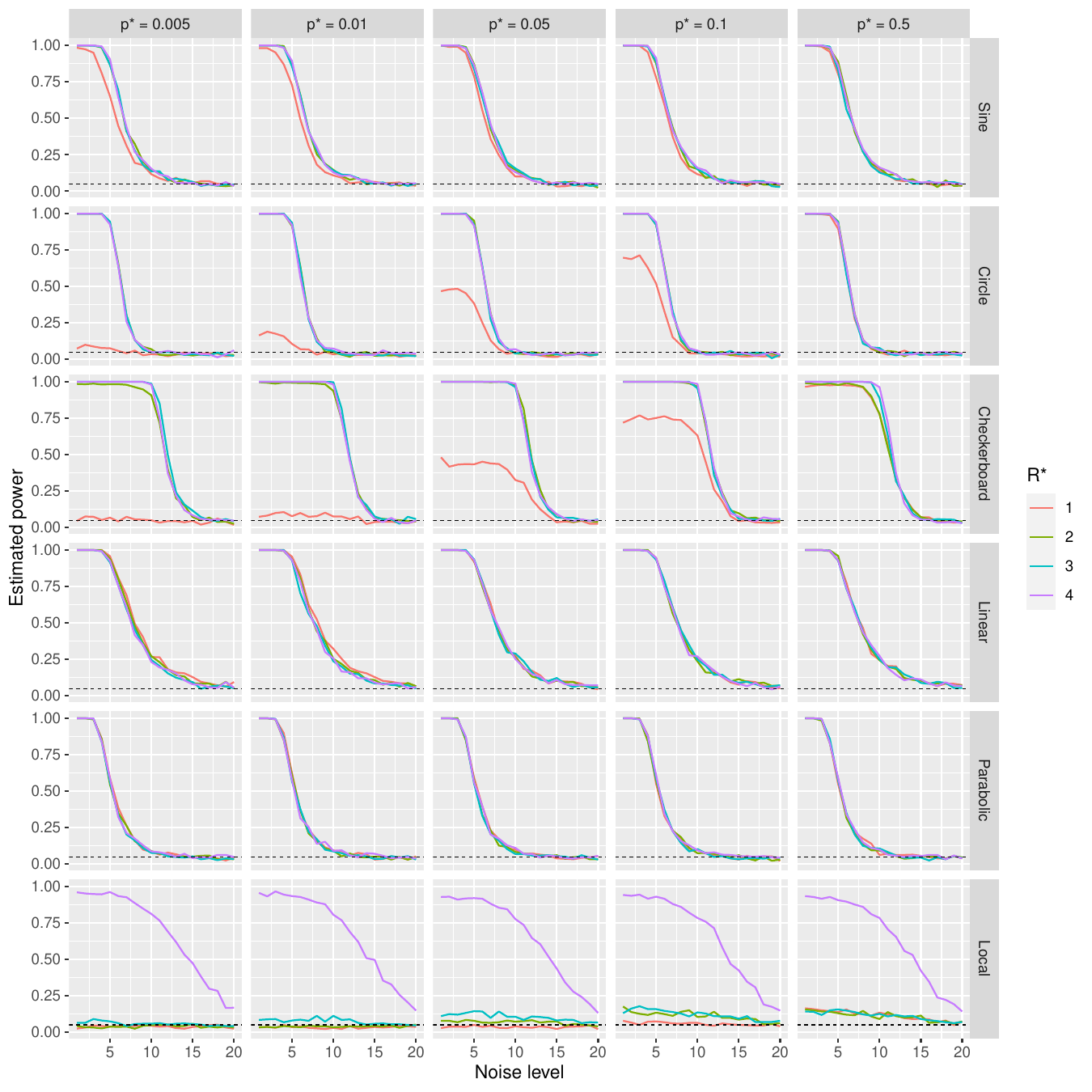}
\caption{Power versus noise level for different methods. Estimated power at 20 noise levels for \texttt{MultiFIT} with different $p^*$ and $R^*$ values under the six scenarios from {\bf Table~\ref{tbl:scenarios}}\label{fig:param_grid_1}.}
\end{figure}

\begin{figure}[ht]
    \centering
    \includegraphics[width=1\textwidth]{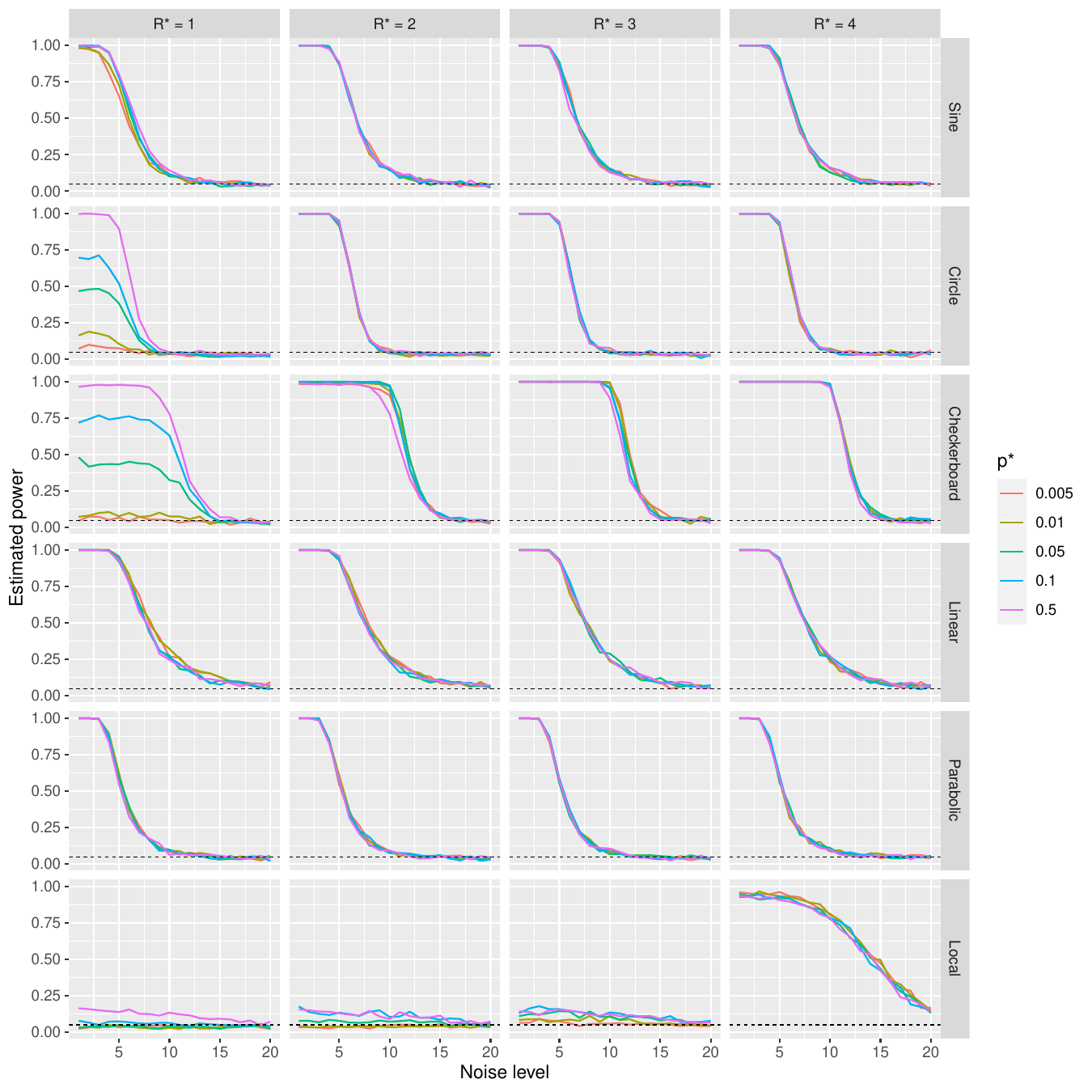}
\caption{Power versus noise level for different methods. Estimated power at 20 noise levels for \texttt{MultiFIT} with different $p^*$ and $R^*$ values under the six scenarios from {\bf Table~\ref{tbl:scenarios}}\label{fig:param_grid_2}.}
\end{figure}

In \textbf{Figures~\ref{fig:param_grid_1}} and \textbf{\ref{fig:param_grid_2}} we demonstrate the effect of our main tuning parameters -- $p^*$ and $R^*$ -- on the estimated power. We evaluate the power for the simulation scenarios that are detailed in \textbf{Table~\ref{tbl:scenarios}} over the grid $\mathbf{p}^*\times \mathbf{R}^*$ where $\mathbf{p}^* = \{0.005, 0.01, 0.05, 0.1, 0.5\}$ and $\mathbf{R}^* = \{1,2,3,4\}$. In general, higher $p^*$ and higher $R^*$ values entail more testing. Under some scenarios, where the dependency structure is ``more local'', higher values are needed to ensure better power. When the dependency structures are ``more global'' we see that the selection of the tuning parameters makes little difference to the estimated power.

\end{document}